\documentclass[onecolumn, journal,11pt]{IEEEtran}

\IEEEoverridecommandlockouts
\title{Secret Key Establishment \\ over a Pair of Independent Broadcast Channels}
\author{Hadi Ahmadi, Reihaneh Safavi-Naini \\
\footnotesize{Department of Computer Science, University of Calgary, Canada.}\\
\footnotesize{\{hahmadi, rei\}@ucalgary.ca}}
\usepackage{multirow}
\usepackage{color}
\usepackage{cite}
\usepackage{epsfig}
\usepackage{graphicx}
\graphicspath{{../figures/}}
\usepackage{subfigure}
\usepackage{dsfont}
\usepackage{amssymb}
\usepackage{amsmath}
\usepackage{enumerate}

\newcommand{\remove}[1]{}

\newcommand{\tab}{\hspace*{4em}}
\newtheorem{theorem}{Theorem}
\newtheorem{corollary}{Corollary}
\newtheorem{lemma}{Lemma}
\newtheorem{definition}{Definition}

\begin{document}
\maketitle

\footnotetext{This work has been submitted to the 2010 International Symposium on Information Theory and its Applications (ISITA2010).}
\renewcommand{\baselinestretch}{1.2}
\normalsize
\begin{abstract}
This paper considers the problem of information-theoretic Secret Key Establishment (SKE) in the presence of a passive adversary, Eve, when Alice and Bob are connected by a pair of independent discrete memoryless broadcast channels in opposite directions. We refer to this setup as \emph{2DMBC}. We define the secret-key capacity in the 2DMBC setup and prove lower and upper bounds on this capacity. The lower bound is achieved by a two-round SKE protocol that uses a two-level coding construction. We show that the lower and the upper bounds coincide in the case of degraded DMBCs.
\end{abstract}

\section{Introduction}
Secret Key Establishment (SKE) is a fundamental problem in cryptography: Alice and Bob want to share a secret key in the presence of an adversary, Eve. We consider information theoretic SKE where there is no assumption on Eve's computational power and assume Eve is passive and can only eavesdrop the communication between Alice and Bob.
It has been proven that SKE is impossible if Alice and Bob are connected by an insecure and reliable channel with no prior correlated information \cite{Ma93}. Thus, information-theoretic solutions to the SKE problem assume that resources such as channels and/or correlated sources are available to the parties. We refer to a specific collection of resources available to the parties as a \emph{setup}.

\begin{figure}[ht]
\centering
\subfigure[Wyner's wire-tap channel]{
\includegraphics[scale=.6]{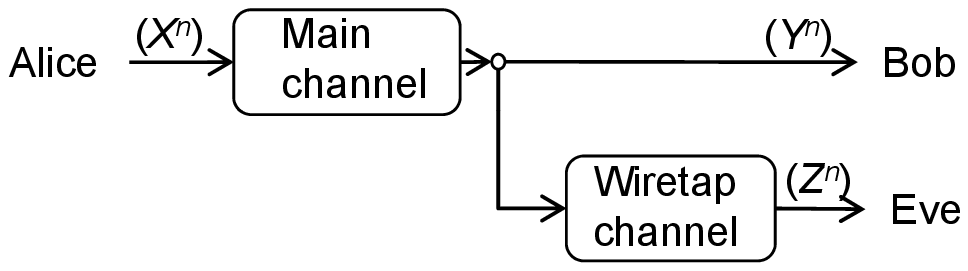}
\label{fig1:subfig1}
}
\subfigure[Csisz$\mathrm{\acute{a}}$r and K$\mathrm{\ddot{o}}$rner's broadcast channel]{
\includegraphics[scale=.6]{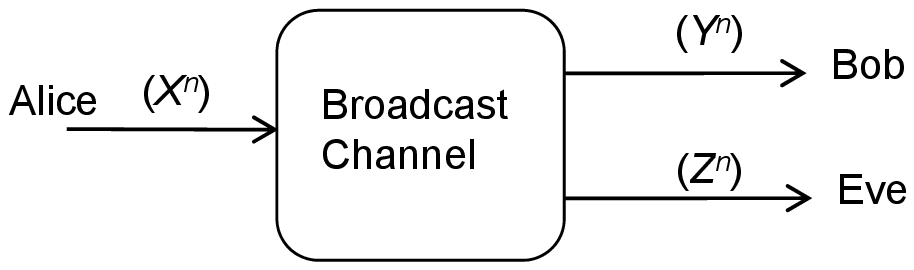}
\label{fig1:subfig2}
}
\caption[Optional caption for list of figures]{The comparison of \subref{fig1:subfig1} Wyner's wire-tap channel and \subref{fig1:subfig2} Csisz$\mathrm{\acute{a}}$r and K$\mathrm{\ddot{o}}$rner's broadcast channel}
\label{fig1}
\end{figure}

One method of establishing a secure key between Alice and Bob is Alice choosing a random key and sending it as a message securely to Bob. This is essentially using a secure message transmission protocol for SKE. In a pioneering work, Wyner \cite{Wy75} considered the scenario of secure communication over noisy channels, where there is a Discrete Memoryless Channel (DMC), called the \emph{main channel} from Alice to Bob, and a second DMC, called the \emph{wiretap channel}, from Bob to Eve, through which Eve can observe a (degraded) noisy version of what Bob receives from Alice. See Fig.~\ref{fig1:subfig1}. Wyner defined the {\em secrecy capacity}, $C_s$, in this setup as the highest rate of secure and reliable message transmission from Alice to Bob. He proved a single-letter characterization for the secrecy capacity that implies the possibility of secure message transmission if the main channel has a non-zero (communication) capacity and the wiretap channel is noisy. Wyner's work on secure message transmission is important because, contrary to the well-known Shannon's model of secure communication \cite{Sh48}, (i) it does not assume any prior shared secret key and, (ii) rather than spending resources to realize noiseless channels, it uses channel noise to provide security. Csisz$\mathrm{\acute{a}}$r and K$\mathrm{\ddot{o}}$rner \cite{Cs78} generalized Wyner's wiretap channel setup by introducing \emph{noisy broadcast channel with two receivers}, where there is a Discrete Memoryless Broadcast Channel (DMBC) with one sender (Alice) and two receivers (Bob and Eve). See Fig.~\ref{fig1:subfig2}. They determined the secrecy capacity of this setup and showed that secure message transmission from Alice to Bob is possible if Bob's channel is \emph{less noisy} \cite{Ko77}, compared to Eve's. The results of this study have been extended to the case of Gaussian channels \cite{Le78}.

The work in \cite{Wy75} and \cite{Cs78} can be used for SKE, inasmuch as achievable rates for secure message transmission and secret key establishment become the same in these setups. Later work has followed two directions: one aiming at applying the SKE results to real-life communication scenarios such as SKE in wireless environments \cite{Bl08}, and the second considering SKE in new setups. Public discussion channel \cite{Ah93,Ma93,Cs00}, secure feedback channel \cite{Ah06}, modulo-additive feedback channel \cite{La08,Te08}, and correlated sources \cite{Kh08,Pr08} are examples of new ingredients to build such setups.

Maurer \cite{Ma93} and concurrently Ahlswede and Csisz$\mathrm{\acute{a}}$r \cite{Ah93} studied SKE when there exists a DMBC from Alice to Bob (and Eve) and a public discussion channel between Alice and Bob that is unlimitedly available to send messages in both directions. This latter channel is reliable but insecure, i.e., Eve can fully eavesdrop the communication. It was shown that SKE in this setup may be possible even in cases where the secrecy capacity of the DMBC is zero. The work in \cite{Ah93,Ma93} also includes the setup where the DMBC is replaced with a Discrete Memoryless Multiple Source (DMMS) between the parties. Csisz$\mathrm{\acute{a}}$r and Narayan \cite{Cs00} studied SKE in a slightly different setup that consists of a DMMS and a limited-rate one-way public channel from Alice to Bob. Ahlswede and Cai \cite{Ah06} showed that the secrecy capacity in Wyner's setup can be increased by adding an unlimited secure (and reliable) \emph{output feedback channel}. This channel is only used to feed back the information received at the output of the forward channel. Noisy feedback over modulo-additive broadcast channels \cite{Te08, La08} is another extension of the SKE problem. SKE using a DMBC from Alice to Bob and (Eve) and a DMMS between the three parties was considered in \cite{Kh08} and independently in \cite{Pr08}.

Assuming the existence of (free) public discussion, secure feedback, or modulo-additive feedback channels lets us build setups that allow interactive communication between Alice and Bob. In these setups, Alice and Bob can benefit from multi-round SKE protocols to achieve higher secret-key rates. In practice, however, such channels may not exist and it may not be necessarily the best strategy (for maximizing the secret-key rate) to realize them from given resources.

\subsection{Our work}
We consider a new setup for SKE where Alice and Bob are connected by a pair of independent DMBCs in opposite directions. We refer to this setup as \emph{2DMBC}. This setup is a realistic scenario that models wireless networks where two nodes communicate over wireless channels in two directions, and their communication is eavesdropped by neighbors in their communication range. The 2DMBC setup gives the promise of interactive communication, while the only resources provided to the parties are DMBCs.

We define SKE in the 2DMBC setup as a multi-round protocol between Alice and Bob with the aim of establishing a secure and reliable key. In analogy to the secrecy capacity \cite{Wy75,Cs78,Ma93}, we define the \emph{secret-key capacity} in this setup, denoted by $C^{2DMBC}_{sk}$, as the maximum achievable secret-key rate, in bits per use of the channel. We have the following results.

\subsubsection{Lower bound} We give a lower bound on the secret-key capacity. We propose a two-round SKE protocol that uses a two-level channel coding construction, and prove that it achieves the lower bound. Our lower bound can also be derived by using the SKE protocols in the DMMS-and-DMBC setup \cite{Kh08,Pr08}. However, while the SKE protocols proposed in \cite{Kh08,Pr08} are combinations of different constructions for different cases (depending on the setup's specification), our proposed SKE protocol uses a concrete construction that achieves the lower bound for all cases.

\subsubsection{Upper bound} We prove an upper bound on the secret-key capacity. This bound holds for all the secret-key rates achievable by SKE protocols with no limitation on the number of communication rounds.

\subsubsection{Degraded 2DMBCs} We study the 2DMBC setup when the broadcast channels are degraded. We show that in this setup the lower and the upper bounds coincide, and the secret-key capacity can be achieved by a one round SKE protocol. This implies that, in the case of degraded 2DMBCs, interactive communication cannot improve the secret-key rate and the optimal solution is \emph{key transport}, i.e., one party choosing a key and sending it securely though the (one-way) DMBC, i.e., following the the work in \cite{Cs78}.

\subsection{Discussion}
\subsubsection{Types of key establishment protocols} \label{sec-types of SKE}
We observe that SKE in the 2DMBC setup can take one of the following forms:
\begin{itemize}
\item[(A)] \textit{Key Transport,} where one party selects the
key prior to the start of the protocol and the protocol is mainly
used to deliver the key to the recipient in a secure and reliable manner.
\item[(B)] \textit{Key Agreement,} where the final secret key
is not selected by a single party prior to the start of the protocol. Instead, it is a (possibly randomized) function of the inputs of the two parties. The randomness in the function comes from the channel noise.
\end{itemize}
We note that method (A) is essentially secure message transmission, while method (B) is purely for sharing a secret key. It may be argued that key agreement protocols (type (B)) offer a higher level of security as the key is not determined by a single party.

\subsubsection{Secrecy capacity vs. secret-key capacity}
The secrecy capacity was originally defined in \cite{Cs78,Wy75} for secure message transmission over one-way noisy channels. The definition secret-key capacity was first defined in \cite{Ah93}. Following these two definitions, one can define secrecy capacity and secret-key capacity for a given setup. As discussed in Section \ref{sec-types of SKE}, a protocol for secure message transmission in a setup can always be used for SKE in that setup, and so, in any setup, \emph{the secret-key capacity is at least equal to the secrecy capacity}.

In \cite{Cs78,Wy75}, there is only a one-way channel from Alice to Bob (and Eve) and the only way to establish a key is to use choose one and send it using a secure message transmission protocol. Hence the secret-key capacity is equal to the secrecy capacity. The same result holds for setups that include a (free) public discussion channel \cite{Ma93,Ah93,Cs00} since any SKE protocol can be used along with a one-time pad encryption for the purpose of secure message transmission. In the 2DMBC setup, however, the two capacities are not necessarily the same. This is because the only accessible channels are noisy channels and to send the encrypted message (using the established key) a reliable communication channel needs to be constructed first. The relationship between the two capacities is not in the scope of this paper.

\subsubsection{Strong and weak secrecy/secret-key capacity}
The notion of secret-key capacity defined in this paper follows the definition of secrecy capacity in \cite{Wy75} and later in \cite{Cs78,Ma93, Ah93, Kh08,Pr08}. The secrecy requirement in these definitions is ``weak'' because it requires Eve's uncertainty rate to be negligible. A ``stronger'' variation is requiring Eve's total uncertainty to be negligible. Maurer and Wolf \cite{Ma00} showed that replacing the (weak) secrecy requirement by the stronger one does not decrease the secrecy capacity of setups considered in \cite{Wy75,Cs78,Ma93}. A similar proof can be used to show that the secrecy-key capacity in the 2DMBC setup remains the same, regardless of which secrecy requirement is used. This means that our results are also valid for the strong secret-key capacity.

\subsection{Notation}
We use calligraphic letters ($\mathcal{U}$) to denote finite alphabets. We denote random variables (RVs) and their realizations over these sets by the corresponding letters in uppercase ($U$)  and lowercase ($u$). The size of the set $\mathcal{U}$ is denoted by $|\mathcal{U}|$. $\mathcal{U}^n$ is the set of all sequences of length $n$ (so called $n$-sequences) with elements from $\mathcal{U}$.  $U^n =(U_1,U_2, \dots, U_n) \in \mathcal{U}^n$ denotes a random $n$-sequence in $\mathcal{U}^n$.

Let $X$ be an RV over the set $\mathcal{X}$, denoted by $X\in \mathcal{X}$. We denote its probability distribution by $P_X$ and its entropy by $H(X)$. Given a pair of RVs, $(X,Y)\in \mathcal{X}\times \mathcal{Y}$, we denote the joint distribution of $X$ and $Y$ by $P_{X,Y}$ and their joint entropy by $H(X,Y)$. The conditional probability distribution and the entropy of $Y$ given $X$ are denoted by $P_{Y|X}$ and $H(Y|X)$, respectively. The mutual information between $X$ and $Y$ is denoted by $I(X;Y)$. Given RVs $(X,Y,Z)\in \mathcal{X}\times \mathcal{Y} \times \mathcal{Z}$, we denote by $P_{Y,Z|X}$ the conditional joint distribution of $Y$ and $Z$ when $X$ is known, and by $I(X;Y|Z)$ the mutual information between $X$ and $Y$ when $Z$ is known. $X\leftrightarrow Y\leftrightarrow Z$ denotes a Markov chain between the RVs $X$, $Y$, and $Z$ in this order. We use `$||$' to show the concatenation of two sequences. For a value $x$, we use $[x]_+$ to show $\max\{0,x\}$.

\subsection{Paper organization}
The rest of the paper is organized as follows. Section \ref{sec_definition} gives the setup and definitions. In Section \ref{sec_lowerbound}, we prove a lower bound on the secret-key capacity in this setup. We prove an upper bound on this capacity in Section \ref{sec_upperbound}. The degraded 2DMBC setup is studied in Section \ref{sec-degraded}. Section \ref{sec_conclusion} gives the concluding remarks.

\section{Preliminaries and Definitions}\label{sec_definition}
A Discrete Memoryless Channel (DMC), denoted by $X\rightarrow Y$, is a channel with input and output alphabet sets $\mathcal{X}$ and $\cal Y$, respectively, where each input symbol $X \in {\cal X}$ to the channel results in a single output symbol $Y \in {\cal Y}$, that is independent of previously communicated symbols. The channel is specified by the conditional distribution $P_{Y|X}$.

A Discrete Memoryless Broadcast Channel (DMBC), denoted by $X\rightarrow (Y,Z)$, consists of two (not necessarily independent) DMC's, i.e., $X\rightarrow Y$ and $X\rightarrow Z$. The channel is specified by the conditional distribution $P_{Y,Z|X}$. The {\em secrecy capacity} of the DMBC, $X\rightarrow (Y,Z)$, is defined as the maximum rate at which Alice can reliably send information to Bob such that the rate at which Eve receives this information is arbitrarily small \cite{Wy75,Cs78}.

\begin{definition}\cite{Wy75,Cs78}
The secrecy capacity of the DMBC, specified by $P_{Y,Z|X}$, is denoted by $C_s(P_{Y,Z|X})$, and is defined as the maximum real number $R_s\geq 0$, such that for
every $\delta>0$ and for sufficiently large $N$, there exists a
(possibly probabilistic) $(2^k,N)$ encoder, $e:\{0,1\}^k\rightarrow
{\cal X}^N$ with a decoder, $d:{\cal Y}^N \rightarrow \{0,1\}^k$,
such that for a uniformly distributed binary $k$-sequence $W^k$, we
have $X^N=e(W^k)$, $W'^k=d(Y^N)$ and the following conditions are
satisfied:
\begin{center}
$\begin{cases}
\frac{k}{N} > R_s-\delta\\
\frac{1}{k}H(W^k|Z^N)>1-\delta \\
\Pr(W'^k\neq W^k) < \delta
\end{cases}$.
\end{center}
\end{definition}

It has been proved that \cite{Cs78}
\begin{equation}\label{Csdef}
C_s(P_{Y,Z|X})=\max_{P_{W,X}} \left[I(W;Y)-I(W;Z)\right] \geq \max_{P_X} [I(X;Y)-I(X;Z)],
\end{equation}
where $W$ is a random variable from an arbitrary set $\mathcal{W}$
such that $W\leftrightarrow X \leftrightarrow (Y,Z)$ forms a Markov chain.

\begin{figure} [h]
\centering
  \includegraphics[scale=.35]{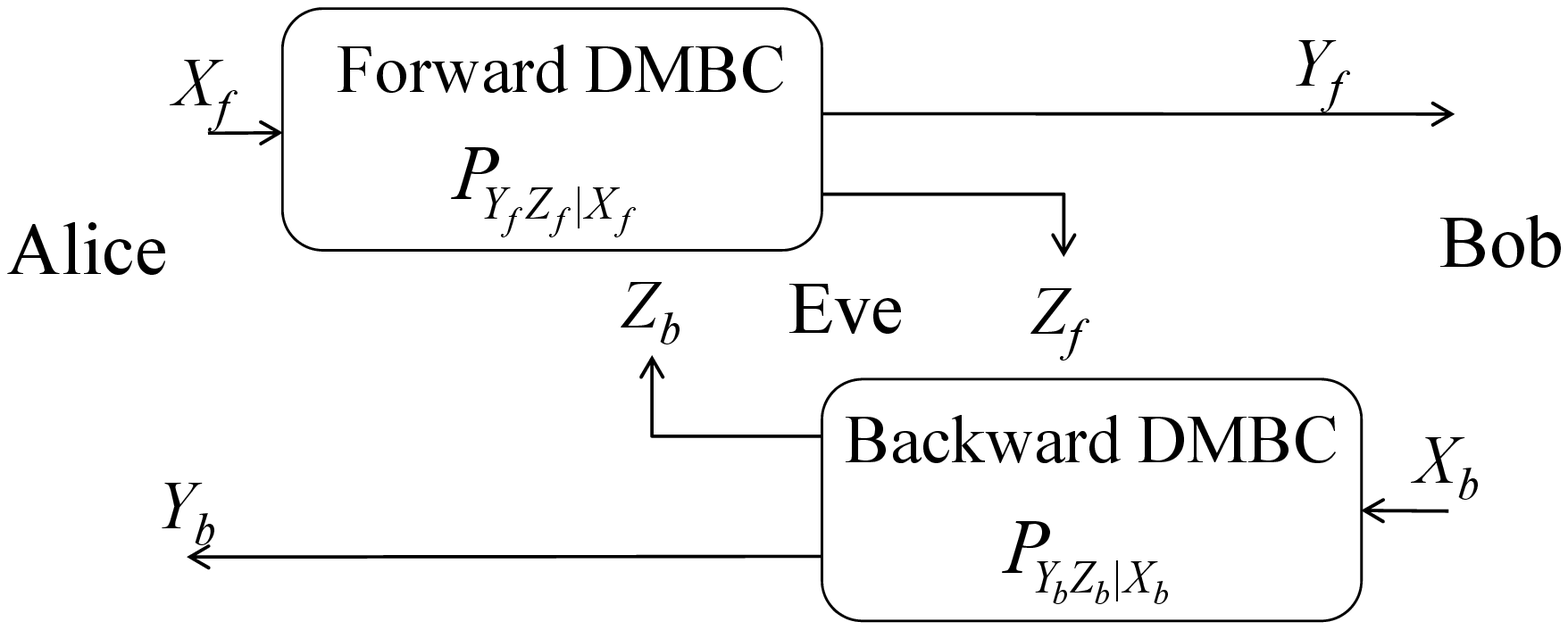}\\
  \caption{The 2DMBC setup}\label{fig_two-way}
\end{figure}

We define a 2DMBC as a pair of independent DMBCs, i.e., a forward DMBC from Alice to Bob, $X_f\rightarrow (Y_f,Z_f)$, specified by $P_{Y_f, Z_f|X_f}$ over the finite sets $\mathcal{X}_f,\mathcal{Y}_f,\mathcal{Z}_f$, and a backward DMBC from Bob to Alice, $X_b\rightarrow (Y_b,Z_b)$, specified by $P_{Y_b, Z_b|X_b}$ over
$\mathcal{X}_b,\mathcal{Y}_b,\mathcal{Z}_b$. See Fig. \ref{fig_two-way}.

We consider the scenario where the 2DMBC is used to establish a shared secret key between Alice and Bob. Alice and Bob use a (possibly) multi-round SKE protocol to exchange sequences of RVs in consecutive rounds. In each communication round, either Alice or Bob sends a sequence of random variables (RVs) as the DMBC input. The legitimate receiver (in this round) computes a sequence of RVs to be sent in the next communication round. This sequence may depend on all previously communicated (sent and/or received) sequences of RVs. At the end of the last communication round, each party (including Eve) will have a set of communicated sequences, which form their ``view'' of the protocol. Let the RVs $View_A$, $View_B$, and $View_E$ be the views of Alice, Bob, and Eve, respectively. Using their views, either Alice or Bob computes a secret key $S$, while the other one computes an estimation of the key $\hat{S}$. In a secure SKE protocol, the established key is required to be \emph{random}, \emph{reliable} and \emph{secret}. These security requirements are formally defined below.

\begin{definition}\label{definition-secure SKE}
For $R_{sk}\geq 0$ and $0\leq \delta \leq 1$, the SKE protocol $\Pi$ in the 2DMBC setup is $(R_{sk}, \delta)$\emph{-secure} if it results in the key $S$ and its estimation $\hat{S}$ such that
\begin{IEEEeqnarray}{l} \label{SKE-Eqs}
\frac{H(S)}{n_f+n_b} >  R_{sk}-\delta,\IEEEyessubnumber \label{SKE-rand} \\
\Pr(\hat{S}\neq S) < \delta,\IEEEyessubnumber \label{SKE-rel} \\
\frac{H(S|View_E)}{H(S)}>1-\delta,\IEEEyessubnumber\label{SKE-sec}
\end{IEEEeqnarray}
where $n_f$ and $n_b$ are the number of times that the forward and the backward channels are used, respectively.
\end{definition}

When $\delta$ tends to zero, $R_{sk}$ indicates the secret-key rate achievable by protocol $\Pi$, i.e., the ratio of the key entropy to the total number of channel uses. We define the secret-key capacity as follows.
\begin{definition}\label{definition-secret-key capacity}
The \emph{secret-key capacity} of a 2DMBC, with forward and backward channels specified by $P_{Y_f, Z_f|X_f},P_{Y_b, Z_b|X_b}$, is denoted by $C^{2DMBC}_{sk}(P_{Y_f, Z_f|X_f},P_{Y_b, Z_b|X_b})$ and is defined as the largest $R_{sk} \geq 0$ such that, for any arbitrarily small $\delta>0$, there exists an $(R_{sk}, \delta)$-secure SKE protocol.
\end{definition}

\section{The Secret-Key Capacity: Lower Bound}\label{sec_lowerbound}
Let the RVs $X_f,Y_f,Z_f$ (resp. $X_b,Y_b,Z_b$) be consistent with the distribution $P_{Y_f,Z_f|X_f}$ (resp. $P_{Y_b,Z_b|X_b}$), specified by the channel. Let $V_{f}$, $V_{b}$, $W_{1,f},W_{2,f}$, $W_{1,b},W_{2,b}$ be random variables from arbitrary sets where, $V_{f}$, $V_{b}$, $ (W_{1,f},W_{2,f})$, and $(W_{1,b},W_{2,b})$ are independent and the following Markov chains are  satisfied:
\begin{subequations}\label{Markov chains1}
\begin{eqnarray}
&&V_{f}\leftrightarrow Y_f \leftrightarrow (X_f,Z_f) \\
&&W_{2,b}\leftrightarrow W_{1,b}\leftrightarrow X_b \leftrightarrow (Y_b,Z_b), \\
&&V_{b} \leftrightarrow Y_b \leftrightarrow (X_b,Z_b) \\
&&W_{2,f}\leftrightarrow W_{1,f}\leftrightarrow X_f \leftrightarrow (Y_f,Z_f).
\end{eqnarray}
\end{subequations}

\begin{theorem} \label{theorem-lowerbound}
Taking the above variables and letting
\begin{subequations} \label{R^AB_s12}
\begin{eqnarray}
&&R^A_{s1}=I(V_{f};X_f)-I(V_{f};Z_f),\label{R_s1^A} \\
&&R^A_{s2}=I(W_{1,b};Y_b|W_{2,b})-I(W_{1,b};Z_b|W_{2,b}), \label{R_s2^A}\\
&&R^B_{s1}=I(V_{b};X_b)-I(V_{b};Z_f),\\
&&R^B_{s2}=I(W_{1,f};Y_f|W_{2,f})-I(W_{1,f};Z_f|W_{2,f}),
\end{eqnarray}
\end{subequations}
the secret-key capacity is lower bounded as
\begin{eqnarray} \label{C^SC}
C^{2DMBC}_{sk} \geq \max \{L_A,L_B\},
\end{eqnarray}
where
\begin{IEEEeqnarray}{ll}
L_A =  \max_{n_f,n_b,P_{X_f,V_{f}},P_{X_b, W_{2,b},W_{1,b}}} \left[  \frac{n_f R^A_{s1}+ n_b [R^A_{s2}]_+}{n_f+n_b} ~\mathrm{s.~t.}~~ n_f I(V_{f};Y_f|X_f) < n_b I(W_{1,b};Y_b)\}\right],  \label{L_A} \\
\nonumber \\
L_B =  \max_{n_f,n_b,P_{X_b,V_{b}}, P_{X_f,W_{2,f},W_{1,f}}} \left[  \frac{n_b R^B_{s1} + n_f [R^B_{s2}]_+}{n_f+n_b} ~\mathrm{s.~t.}~~ n_b I(V_{b};Y_b|X_b) < n_f I(W_{1,f};Y_f)\}\right]. \label{L_B}
\end{IEEEeqnarray}
\end{theorem}
\emph{Proof}: Appendix \ref{app_A}.

The proof of Theorem \ref{theorem-lowerbound} uses a concrete two-round SKE protocol with a two-level coding construction. We give an outline of the protocol for a special case where Alice is the initiator, and we have $V_f=Y_f$, $W_{1,b}=X_b$, and $W_{2,b}=1$. Let $\eta_f$, $\eta_t$, $\eta_b$, $R^A_s$, and $\kappa$ be defined as
\begin{eqnarray*}
&&\eta_f = n_f H(Y_f),\\
&&\eta_t = n_f H(Y_f|X_f),\\
&&\eta_b = n_b I(X_b;Y_b)- \eta_t,\\
&&R_s^A=\frac{n_f R^A_{s1} + n_b [R^A_{s2}]_+}{n_f+n_b},\\
&&\kappa = (n_f+n_b) R_s^A.
\end{eqnarray*}

Alice chooses $n_f$ copies of $X_f$ independently and identically distributed (i.i.d.) w.r.t. $P_{X_f}$ to create the $n_f$-sequence $X^{n_f}_f$, and sends it over the forward DMBC. Bob receives $Y^{n_f}_f$ and maps it to an integer $F\in \mathcal{F}=\{1,2,\dots,2^{\eta_f}\}$ using a deterministic bijective mapping. He encodes $F$ to an integer $T \in \mathcal{T}=\{1,2,...,2^{\eta_t}\}$; this is the first level of encoding. Bob chooses a uniformly random integer $B \in \mathcal{B}=\{1,2, \dots, 2^{\eta_b} \}$ and encodes $(T,B)$ to an $n_b$-sequence $X^{n_b}_b$; this is the second level of encoding. The constructions of these encoders for the general case are described in Appendix \ref{app_A}. Bob sends $X^{n_b}_b$ over the backward channel and Alice receives $Y^{n_b}_b$. She first decodes $Y^{n_b}_b$ to $(\hat{T},\hat{B})$ and then uses $\hat{T}$ to find the appropriate codebook for decoding $X^{n_f}_f$ and to $\hat{F}$ (and hence $\hat{Y}^{n_f}_f$). The decoder uses the jointly-typical decoding technique.

The secret key is obtained by calculating $S=g(F,B)$, where $g$ is a function defined as follows. Letting $\{\mathcal{G}_i\}_{i=1}^{2^\kappa}$ be a partition of $\mathcal{F} \times \mathcal{B}$ into $2^\kappa$ equal-sized parts, the function $g:\mathcal{F} \times \mathcal{B} \rightarrow \{1,2,\dots,2^\kappa\}$ is such that, for every input $F,B \in \mathcal{G}_i$, outputs $i$. In Appendix \ref{app_A}, we show that there exist appropriate encoding and decoding functions that can be used to achieve the lower bound.

\section{The Secret-Key Capacity: Upper Bound}\label{sec_upperbound}
Let the RVs $X_f,Y_f,Z_f$ and $X_b,Y_b,Z_b$ correspond to the 2DMBC setup specified by $P_{Y_f,Z_f|X_f}$ and $P_{Y_b,Z_b|X_b}$, respectively.

\begin{theorem}\label{th_upperbound}
The secret-key capacity in the 2DMBC setup is upper bounded as
\begin{eqnarray} \label{upper-bound}
C^{2DMBC}_{sk} \leq \max_{P_{X_f},P_{X_b}} \{ I(X_f;Y_f|Z_f) , I(X_b;Y_b|Z_b) \}
\end{eqnarray}
\end{theorem}
\emph{Proof}: Appendix \ref{app_B}.

The upper bound is proved for the highest key rate achievable by a general SKE protocol with an arbitrary number of communication rounds.

\section{Degraded 2DMBCs}\label{sec-degraded}
We define degraded 2DMBCs and prove that the lower and the upper bounds on $C^2_s$ coincide in the case of degraded 2DMBCs. Moreover, this capacity is achieved by a one-round SKE protocol that uses one of the DMBCs.

\begin{definition}\label{O/R-DMBC}
The DMBC $X\rightarrow (Y,Z)$ is called \emph{obversely degraded} if $X \leftrightarrow Y \leftrightarrow Z$ forms a Markov chain. It is called \emph{reversely degraded} if $X \leftrightarrow Z \leftrightarrow Y$ forms a Markov chain.
\end{definition}

We say the DMBC $X\rightarrow (Y,Z)$ \emph{has two independent subchannels}, $X_O\rightarrow (Y_O,Z_O)$ and $X_R\rightarrow (Y_R,Z_R)$, if its input $X$ and output $(Y,Z)$ can be represented as $X=[X_O,X_R]$, $Y=[Y_O,Y_R]$ and $Z=[Z_O,Z_R]$, respectively, such that
\[ (Y_O,Z_O) \leftrightarrow X_O \leftrightarrow X_R \leftrightarrow (Y_R,Z_R) \]
forms a Markov chain.

\begin{definition}\label{deg-DMBC}
The DMBC $X\rightarrow (Y,Z)$ is called degraded if it can be represented by two independent subchannels, $X_O\rightarrow (Y_O,Z_O)$ and $X_R\rightarrow (Y_R,Z_R)$, such that the former channel is obversely degraded and the latter channel is reversely degraded, implying that
\begin{IEEEeqnarray*}{l}
Z_O \leftrightarrow Y_O \leftrightarrow X_O \leftrightarrow X_R \leftrightarrow Z_R \leftrightarrow Y_R
\end{IEEEeqnarray*}
is a Markov chain.
\end{definition}
Note that Definition \ref{deg-DMBC} covers cases where the DMBC is either obversely or reversely degraded. In such cases, in fact only one of the subchannels exists, and the other one can be defined over empty sets of input and outputs.

\begin{definition}\label{O/R-2DMBC}
A 2DMBC is called degraded if both of its one-way DMBCs are degraded.
\end{definition}

\begin{theorem}\label{th_degraded}
For the degraded 2DMBC, specified by $X_f\rightarrow (Y_f,Z_f)$ and $X_b\rightarrow (Y_b,Z_b)$, where
\begin{IEEEeqnarray*}{c}
X_f=[X_{f,O},X_{f,R}],~ Y_f=[Y_{f,O},Y_{f,R}],~ Z_f=[Z_{f,O},Z_{f,R}],\\
X_b=[X_{b,O},X_{b,R}],~ Y_b=[Y_{b,O},Y_{b,R}],~ Z_b=[Z_{b,O},Z_{b,R}],
\end{IEEEeqnarray*}
we have
\begin{IEEEeqnarray*}{l} \label{coincide}
C^{d-2DMBC}_{sk} = \max_{P_{X_{f,O},X_{b,O}}} \{ I(X_{f,O};Y_{f,O}|Z_{f,O}), I(X_{b,O};Y_{b,O}|Z_{b,O}) \}.
\end{IEEEeqnarray*}
\end{theorem}
\emph{Proof}: Appendix \ref{app_C}.

\section{Conclusion}\label{sec_conclusion}
The work on key establishment over a pair of independent discrete broadcast channels (the 2DMBC setup) is inspired by real-life communication between peers, e.g., in wireless environments where the communication between two peers is intercepted by neighbors in the communication range. We defined the secret-key capacity in this setup and provided lower and upper bounds on this capacity. The lower bound is achieved by a two-round SKE protocol that uses a two-level coding construction. We showed that, when the broadcast channels are degraded, the lower and the upper bounds coincide and the secret-key capacity is achieved by a one-round SKE protocol using one of the DMBCs.

\newpage
\appendices

\section{Proof of Theorem \ref{theorem-lowerbound}, the lower bound}\label{app_A}
In parts of the proof, we use the channel coding theorem (e.g., \cite[Theorem 8.7.1]{Co06}), with a decoding method based on so called \emph{jointly-typical bipartite sequences}. A bipartite sequence $X^N=(U^n||T^d)$ is the concatenation of two subsequences, $U^n$ and $T^d$, with two (possibly different) probability distributions, $P_{U^n}$ and $P_{T^d}$, respectively, where $N=n+d$. We extend the definitions of jointly typical sequences to bipartite jointly typical sequences as follows.

\begin{definition}
A sequence $x^N=(u^n||t^d)$ is an \emph{$(\epsilon, n)$-bipartite typical sequence} with respect to the probability distribution
pair $(P_U(u), P_T(t))$, iff
\begin{eqnarray} \label{def-x-typical}
|-\frac{1}{N}\log P(x^N)-\frac{nH(U)+dH(T)}{N}|<\epsilon,
\end{eqnarray}
where $P(x^N)$ is calculated as
\begin{eqnarray}
\displaystyle P(x^N)=\prod_{i=1}^N P(x_i)=\prod_{i=1}^n
P_U(u_i)\times \prod_{i=1}^d P_T(t_i).
\end{eqnarray}
\end{definition}
\medskip

\begin{definition} \label{defbip_joi_typ}
A pair of sequences $(x^N,y^N)=((u^n||t^d),(u'^n||t'^d))$ is an
\emph{$(\epsilon, n)$-bipartite jointly typical pair of sequences} with
respect to the probability distribution pair $(P_{U,U'}(u,u'),
P_{T,T'}(t,t'))$, iff $x^N$ and $y^N$ are $(\epsilon, n)$-bipartite
typical sequences with respect to the marginal probability
distribution pairs $(P_{U}(u), P_T(t))$ and $(P_{U'}(u'),
P_T'(t'))$, respectively, and
\begin{eqnarray}\label{def-xy-typical}
|-\frac{1}{N}\log P(x^N,y^N)-\frac{nH(U,U')+dH(T,T')}{N}|<\epsilon,
\end{eqnarray}
where $P(x^N,y^N)$ is calculated as
\begin{eqnarray}
\displaystyle P(x^N,y^N)=\prod_{i=1}^N P(x_i,y_i)=\prod_{i=1}^n
P_{U,U'}(u_i,u'_i)\times \prod_{i=1}^d P_{T,T'}(t_i,t'_i).
\end{eqnarray}
\end{definition}
\medskip

\begin{definition}
The set $A_\epsilon^{(N,n)}$ is the set of all $(\epsilon,
n)$-bipartite jointly typical pairs of sequences
$(x^N,y^N)=((u^n||t^d),(u'^n||t'^d))$ with respect to the
probability distribution pair $(P_{U,U'}(u,u'), P_{T,T'}(t,t'))$.
\end{definition}
\medskip

\begin{theorem} \textbf{(Joint AEP for bipartite sequences)}\label{theorem-AEP}
Let $(X^N,Y^N)=((U^n||T^d),(U'^n||T'^d))$ be a pair of bipartite
random sequences of length $N$, (each part) drawn i.i.d. according
to the distribution pair $(P_{U,U'}(u,u')$, $P_{T,T'}(t,t'))$. Then,
for large enough $n$ and $d$, we have
\begin{enumerate}
  \item $\Pr((X^N,Y^N)\in A_\epsilon^{(N,n)})\rightarrow 1$
  \item $(1-\epsilon)2^{nH(U,U')+dH(T,T')-N\epsilon} \leq |A_\epsilon^{(N,n)}|\leq 2^{nH(U,U')+dH(T,T')+N\epsilon}$
  \item If $\tilde{X}^N$ and $\tilde{Y}^N$ are independent with the same marginal distributions as $P(x^N,y^N)$, i.e., $(\tilde{X}^N,\tilde{Y}^N)$ is generated according to the distribution $P(x^N)P(y^N)$, then
\begin{eqnarray}\label{eqt10}
\displaystyle \Pr((\tilde{X}^N,\tilde{Y}^N)\in
A_\epsilon^{(N,n)})\leq 2^{-nI(U;U')-dI(T;T')+3N\epsilon}.
\end{eqnarray}
\begin{eqnarray}\label{eqt11}
\displaystyle \Pr((\tilde{X}^N,\tilde{Y}^N)\in
A_\epsilon^{(N,n)})\geq
(1-\epsilon)2^{-nI(U;U')-dI(T;T')-3N\epsilon}.
\end{eqnarray}
\end{enumerate}
\end{theorem}
\emph{Proof}: Appendix \ref{app_E}.
\\

To prove Theorem \ref{theorem-lowerbound}, in the following, we propose a two-round SKE protocol, based on a two-level coding construction, that achieves (\ref{L_A}) when Alice initiates the protocol. One can show in a similar way that (\ref{L_B}) is achievable when Bob is the initiator.

Let the RVs $V_{f}, X_f, Y_f, Z_f$, and $W_{1,b}, W_{2,b}, X_b, Y_b, Z_b$ be the same as defined in Section \ref{sec_lowerbound} (for Theorem \ref{theorem-lowerbound}); hence, the Markov chains in (\ref{Markov chains1}) are satisfied. Also let $n_f$ and $n_b$ be integers that satisfy the constraint condition in (\ref{L_A}). For simplicity, we use $W_{1}, W_{2}$, and $V$ to refer to $W_{1,b}, W_{2,b}$, and $V_{f}$, respectively. Accordingly, we write the argument to be maximized in (\ref{L_A}) as
\begin{eqnarray}\label{R_s}
R_{sk}=\frac{n_f R^A_{s1} + n_b [R^A_{s2}]_+ }{n_f+n_b}
\end{eqnarray}
where
\begin{IEEEeqnarray}{l}
R^A_{s1}=I(V;X_f)-I(V;Z_f), \IEEEyessubnumber \label{R^A_s1} \\
R^A_{s2}=I(W_{1};Y_b|W_{2})-I(W_{1};Z_b|W_{2}),\IEEEyessubnumber \label{R^A_s2}
\end{IEEEeqnarray}
and we rephrase the constraint condition in (\ref{L_A}) as
\begin{eqnarray}
n_b I(W_{1};Y_b) \geq n_f (I(V;Y_f|X_f)+ 3\alpha), \label{n_b-n_f}
\end{eqnarray}
where $\alpha>0$ is an small constant to be determined (later) from $\delta$. We shall show that for any given $\delta > 0$, for sufficiently large $n_f$ and $n_b$ that satisfy (\ref{n_b-n_f}), we have
\begin{subequations}
\begin{equation}
\label{uni} \frac{1}{n_f+n_b}H(S) \geq R_s-\delta,
\end{equation}
\begin{equation}
\label{rel} \Pr(\hat{S}\neq S) < \delta,
\end{equation}
\begin{equation}
\label{sec} \frac{H(S|Z^{n_f}_f,Z^{n_b}_b)}{H(S)}>1-\delta.
\end{equation}
\end{subequations}
We describe a two-level coding construction and prove that it can achieve the above secret-key rate. Let $N=n_f+n_b$ and $\epsilon, \beta >0$ be small constants determined from $\alpha$ such that $3 N \epsilon < n_b \beta = n_f \alpha$. Let $n_b=n_{b,1}+n_{b,2}$, where $n_{b,2}$ is chosen to satisfy
\begin{equation} \label{n_{b,2}}
    n_{b,2} I(W_{1};Y_b) =  n_f (I(V;Y_f|X_f)+ 3\alpha).
\end{equation}
We first define the following quantities, sets and function which are used in the sequel.
\begin{eqnarray}
&& \eta_f=n_f [I(V;Y_f)+\alpha], \label{eta_f-def}\\
&& \eta_t=n_{b,2} [I(W_{1};Y_b)-\beta],~~ \eta_{t,2}=n_{b,2} I(W_2;Y_b),~~~ \eta_{t,1}=\eta_t-\eta_{t,2}, \label{eta_t-def}\\
&& \eta_b=n_{b,1} [I(W_{1};Y_b) - \beta],~~  \eta_{b,2}=n_{b,1} I(W_{2};Y_b),~~~ \eta_{b,1}=\eta_b-\eta_{b,2}, \label{eta_b-def}\\
&& \eta_1=\eta_{t,1} + \eta_{b,1},~~ \eta_2=\eta_{t,2} + \eta_{b,2},~~ \eta=\eta_f + \eta_b, \label{eta_1-def}\\
&& \kappa=(n_f+n_b) R_{sk}, ~~ \gamma=\eta-\kappa. \label{kappa&gamma-def}
\end{eqnarray}

Although the quantities obtained in (\ref{n_{b,2}})-(\ref{kappa&gamma-def}) are real values, for sufficiently large $n_b$ and $n_f$, we can approximate them by integers. Since $\beta$ can be made arbitrarily small, we can assume $\eta_b$ and $\eta_t$ are non-negative. Furthermore, it is easy to see that, for arbitrarily small $\alpha$, we can assume $\eta_f\geq \eta_t$ and $\gamma$ is non-negative. We show them respectively as follows
\begin{IEEEeqnarray*}{ll}
\eta_f  &\stackrel{(a)}{=} n_f [I(V;Y_f,X_f)+\alpha]= n_f I(V;X_f)+ n_f I(V;Y_f|X_f)+ n_f \alpha \\
        &= n_f I(V;X_f) + n_{b,2} I(W_1,Y_b) - 2 n_f \alpha \geq n_{b,2} I(W_1,Y_b) - 2 n_f \alpha \geq \eta_t - 2 n_f \alpha,
\\
\\
\eta    & = \eta_f + \eta_b = n_f I(V;X_f) + n_{b,2} I(W_1,Y_b) - 2 n_f \alpha +n_{b,1} I(W_1,Y_b) \\
        & \geq n_f I(V;X_f) + n_{b} I(W_1,Y_b) - 2 n_f \alpha \geq R^A_{s1} + R^A_{s2} -2 n_f \alpha \geq \kappa-2 n_f \alpha,
\end{IEEEeqnarray*}
where equality (a) is due to the Markov chain $X_f\leftrightarrow Y_f \leftrightarrow V$, and the rest of the steps follow from the above relations (\ref{R_s})-(\ref{kappa&gamma-def}). The following sets and functions are used to design the SKE protocol.
\begin{enumerate}[(i)]
    \item $\mathcal{V}^{n_f}$ is the set of all possible $n_f$-sequences with elements from $\mathcal{V}$. Create $\mathcal{V}_{\epsilon}^{n_f}$ by randomly and independently selecting $2^{\eta_f}$ $\epsilon$-typical sequences (w.r.t. $P_V$) from $\mathcal{V}^{n_f}$.
    \item Let $\mathfrak{f}:\mathcal{V}_{\epsilon}^{n_f} \rightarrow \mathcal{F}=\{1,2,\dots,2^{\eta_f}\}$ be an arbitrary bijective mapping; denote its inverse by $\mathfrak{f}^{-1}$.
    \item Let $\{\mathcal{V}_{i,\epsilon}^{n_f}\}_{i=1}^{2^{\eta_t}}$ be a partition of $\mathcal{V}_{\epsilon}^{n_f}$ into $2^{\eta_t}$ equal-sized parts. Define the function $\mathfrak{t}: \mathcal{V}_{\epsilon}^{n_f} \rightarrow \mathcal{T} = \{1,2,\dots,2^{\eta_t}\}$ such that, for any input in $\mathcal{V}_{i,\epsilon}^{n_f}$, it outputs $i$.
    \item Let $\{\mathcal{T}_i\}_{i=1}^{2^{\eta_{t,2}}}$ be a partition of $\mathcal{T}$ into $2^{\eta_{t,2}}$ equal-sized parts. Label elements of part $i$ as $\mathcal{T}_i=\{t_{i,j}\}_{j=1}^{\eta_{t,1}}$. Define $\mathfrak{t}_{indx}:\mathcal{T}\rightarrow \{1,\dots,2^{\eta_{t,2}}\} \times \{1,\dots,2^{\eta_{t,1}}\}$ such that $\mathfrak{t}_{indx}(t)=(i,j)$, if $t$ is labeled by $t_{i,j}$.
    \item Let $\mathcal{B}=\{1,2,\dots,2^{\eta_b}\}$. In analogy to $\mathcal{T}$, let $\{\mathcal{B}_i\}_{i=1}^{2^{\eta_{b,2}}}$ be a partition of $\mathcal{B}$ where $\mathcal{B}_i=\{b_{i,j}\}_{j=1}^{2^{\eta_{b,1}}}$. Define $\mathfrak{b}_{indx}:\mathcal{B}\rightarrow \{1,\dots,2^{\eta_{b,2}}\} \times \{1,\dots,2^{\eta_{b,1}}\}$ such that $\mathfrak{b}_{indx}(b)=(i,j)$, if $b$ is labeled by $b_{i,j}$.
    \item Let $\{\mathcal{G}_i\}_{i=1}^{2^\kappa}$ be a partition of $\mathcal{F} \times \mathcal{B}$ into parts of size $2^{\gamma}$. Define $g:\mathcal{F} \times \mathcal{B} \rightarrow \{1,2,\dots,2^\kappa\}$ such that, for any input in $\mathcal{G}_i$, it outputs $i$.
    \item Define the codebook $\mathcal{C}_{2}$ as a the collection of $2^{\eta_2}$ codewords $\{w^{n_{b}}_{2,t_2,b_2}:~t_2=1,2,\dots,2^{\eta_{t,2}}, ~b_2=1,2,\dots,2^{\eta_{b,2}}\}$, where each codeword $w^{n_{b}}_{2,t_2,b_2}$ is of length $n_{b}$ and is independently generated according to the distribution
         \[ \prod_{i=1}^{n_{b}} p(W_2=w_{2,t_2,b_2}(i)). \]
    \item For each $w^{n_{b}}_{2,t_2,b_2}$, define the codebook $\mathcal{C}_1(w^{n_{b}}_{2,t_2,b_2})$ as the collection of $2^{\eta_1}$ codewords $\{w^{n_{b}}_{1,t_2,b_2,t_1,b_1}:~t_1=1,\dots,2^{\eta_{t,1}} ~b_1=1,\dots,2^{\eta_{b,1}}\}$, where each codeword, $w_{1,t_2,b_2,t_1,b_1}$, is of length $n_{b}$ and is independently generated according to the distribution
        \[\prod_{i=1}^{n_{b}} p(W_1=w_{1,t_2,b_2,t_1,b_1}(i)|W_2=w_{2,t_2,b_2}(i)).\]
    \item Let $Enc:\mathcal{T}\times \mathcal{B} \rightarrow \mathcal{W}_1^{n_b}$ be an encoding function such that $Enc(t,b)=w^{n_{b}}_{1,t_2,b_2,t_1,b_1}$, using the above codebooks, where $(t_2,t_1)=\mathfrak{t}_{indx}(t)$ and $(b_2,b_1)=\mathfrak{b}_{indx}(b)$.
    \item Let $DMC_W$ be the DMC, $W_1\rightarrow X_b$, that is specified by $P_{X_b|W_1}$.
\end{enumerate}

\bigskip

\noindent \textbf{Encoding.} Alice selects an i.i.d. $n_f$-sequence $X^{n_f}_f$ and sends it over the forward DMBC. Bob and Eve receive $Y_f^{n_f}$ and $Z_f^{n_f}$, respectively. Bob finds a sequence $V^{n_f} \in \mathcal{V}_{\epsilon}^{n_f}$ that is $\epsilon$-jointly typical with $Y_f^{n_f}$ (w.r.t. $P_{V,Y_f}$); he returns a NULL if no such sequence is found. He computes $T=\mathfrak{t}(V^{n_f})$ and then selects an independent uniformly random $B \in \mathcal{B}$. Bob calculates $(T_2,T_1)=\mathfrak{t}_{indx}(T)$ and $(B_2,B_1)=\mathfrak{b}_{indx}(B)$, and use them to calculate $W_1^{n_b}=Enc(T,B)$ (see the encoder construction in (ix)). Next, he inputs $W_1^{n_{b}}$ to $DMC_W$ to compute $X^{n_{b}}_b$, and sends $X^{n_{b}}_b$ over the backward DMBC. Alice and Eve receive $Y_b^{n_{b}}$ and $Z^{n_{b}}_b$, respectively.
\medskip

\noindent \textbf{Decoding.} Alice first finds a \emph{unique} codeword $\hat{W}_1^{n_b}\in \mathcal{C}_1$ that is $\epsilon$-jointly typical to $Y^{n_b}_b$ (w.r.t. $P_{W_1,Y_b}$); she returns a NULL if no such sequence is found. She obtains $(\hat{T},\hat{B})$ such that $Enc(\hat{T},\hat{B})=\hat{W}_1^{n_b}$, and then finds a \emph{unique} codeword $\hat{V}^{n_f}\in \mathcal{V}_{\hat{T},\epsilon}^{n_f}$ that is $\epsilon$-jointly typical to $X^{n_f}_f$ (w.r.t. $P_{V,X_f}$); she returns a NULL if no such sequence is found.
\medskip

\noindent \textbf{Key Derivation.} Bob computes $F=\mathfrak{f}(V^{n_f})$ and $S=g(F,B)$; Alice computes $\hat{F}=\mathfrak{f}(\hat{V}^{n_f})$ and $\hat{S}=g(\hat{F},\hat{B})$.
\medskip

\begin{figure}[ht]
\centering
\subfigure[Encoding and decoding]{
\includegraphics[scale=.4]{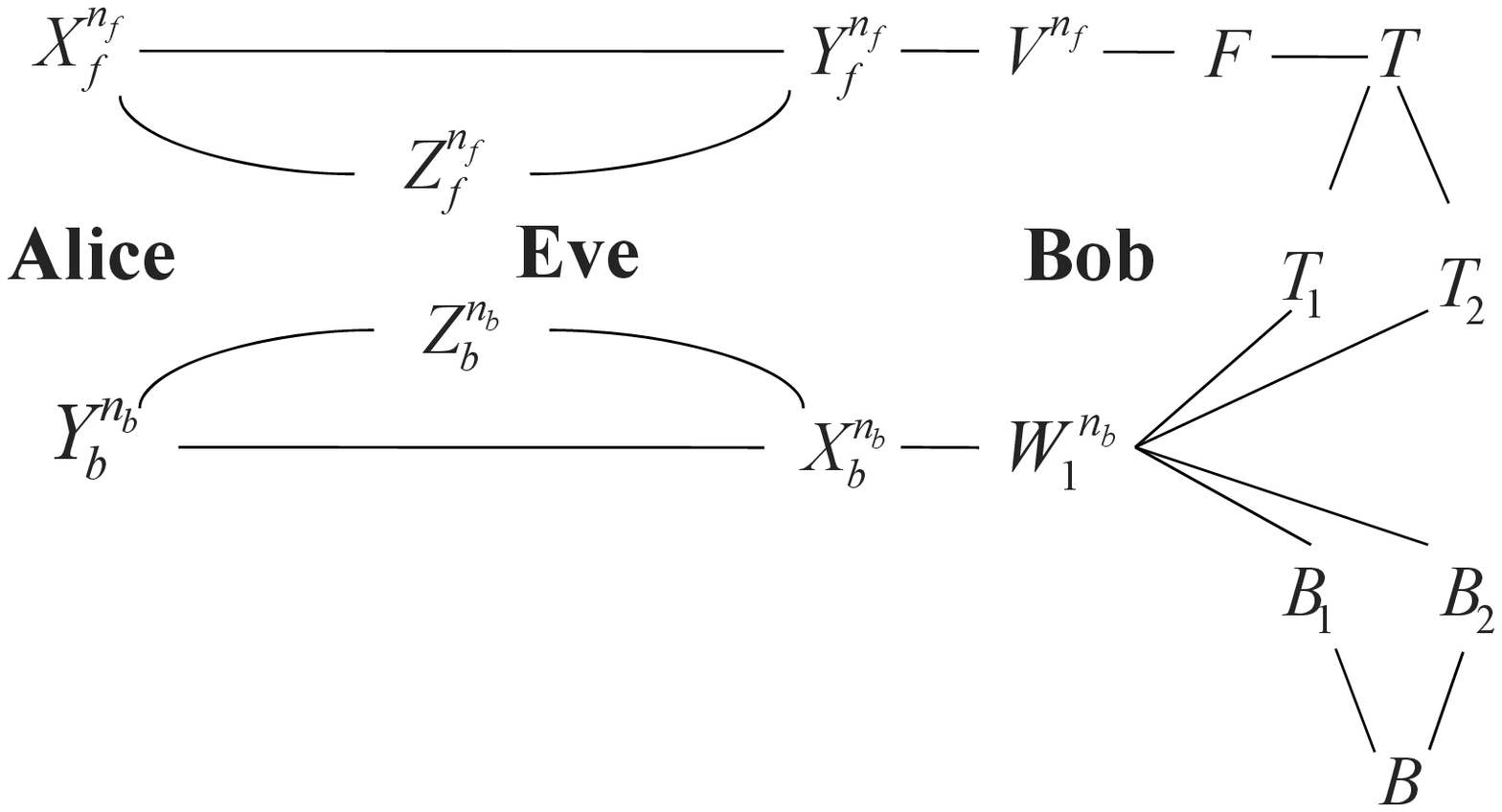}
\label{codec}
}
\subfigure[Key derivation by Alice]{
\includegraphics[scale=.4]{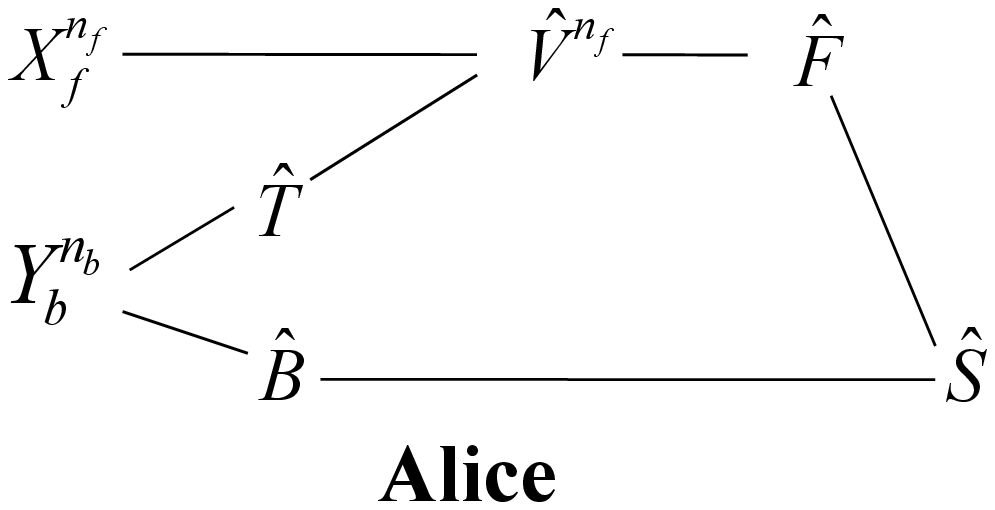}
\label{key_estimate}
}
\subfigure[Key derivation by Bob]{
\includegraphics[scale=.4]{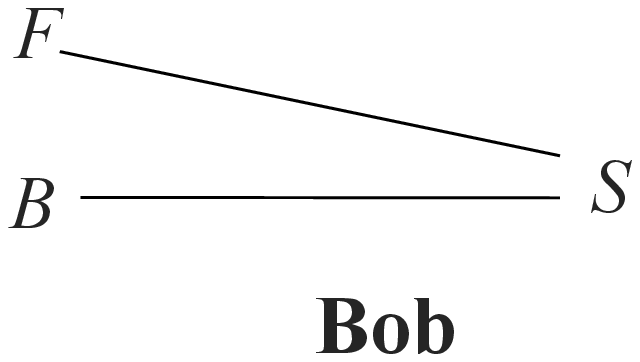}
\label{key_derivate}
}
\caption[Optional caption for list of figures]{The relation between the variables/sequences used in the SKE protocol for \subref{codec} encoding/decoding, \subref{key_estimate} key derivation by Alice, and \subref{key_derivate} key derivation by Bob}
\label{figSKE}
\end{figure}

Fig. \ref{figSKE} shows the connection chain between the random variables/sequences used in the above protocol. Two variables/sequences are connected by an edge if (1) they belong to input/outputs of the same DMBC, or (2) one is computed from the other by Alice or Bob using a (possibly randomized) function. The Markov chain $Q_1 \leftrightarrow Q_2 \leftrightarrow Q_3$ holds, if $Q_3$ (resp. $Q_1$) is computed from $Q_2$ by a (possibly randomized) function $\phi(R,Q_2)$ where $R$ is independent of $Q_1$ (resp. $Q_3$).

\noindent \textbf{Uniformity Analysis: Proving (\ref{uni})}\\
We show that $S \in \{1,2,\dots, 2^{\kappa}\}$ has a distribution close to uniform. First, we argue about the distributions of $V^{n_f}$, $F$, and $B$.

In the encoding phase, $V^{n_f}$ is chosen to be $\epsilon$-jointly typical with $Y_f^{n_f}$ (w.r.t. $P_{V,Y_f}$). From AEP, for each $v^{n_f} \in \mathcal{V}^{n_f}_{\epsilon}$, there are at most $2^{{n_f}(H(Y_f|V)+\epsilon)}$ sequences in $\mathcal{Y}_f^{n_f}$ that are $\epsilon$-jointly typical with $v^{n_f}$; each appearing with probability at most $2^{-{n_f}(H(Y_f)-\epsilon)}$, and so letting
\begin{eqnarray*}
\mathcal{D}_{v^{n_f}}=\{y_f^{n_f}\in \mathcal{Y}_f^{n_f} :~~ (y_f^{n_f},v^{n_f}) \mbox{ is }\epsilon\mbox{-jointly typical w.r.t. }P_{Y_f,V}\},
\end{eqnarray*}
we have
\begin{eqnarray}
\forall v^{n_f} \in \mathcal{V}^{n_f}_{\epsilon},~ \Pr(V^{n_f}=v^{n_f}) &=& \sum_{\mathcal{D}_{v^{n_f}}} \Pr(Y_f^{n_f}=y^{n_f}_f) \Pr(V^{n_f}=v^{n_f}|Y_f^{n_f}=y^{n_f}_f) \nonumber \\
& \leq& \sum_{\mathcal{D}_{v^{n_f}}} \Pr(Y_f^{n_f}=y^{n_f}_f) \nonumber \\
& \leq& 2^{n_f H(Y_f|V)+{n_f}\epsilon} \times 2^{- n_f H(Y_f)+{n_f}\epsilon} \nonumber \\
& =& 2^{{n_f} (-I(V;Y_f)+2\epsilon)} < 2^{-\eta_f + 5 N \epsilon}. \label{P_V}\\
\nonumber \\
\Rightarrow \eta_f - 5 N \epsilon < n_f (I(V;Y_f) - 2 \epsilon) &\leq& H(V^{n_f}) \leq \eta_f = n_f (I(V;Y_f) + \alpha), \label{H_V}
\end{eqnarray}
where the upper bound on $H(V^{n_f})$ is due to $|\mathcal{V}^{n_f}_\epsilon|=2^{\eta_f}$ (see (i)). Since $F=\mathfrak{f}(V^{n_f})$ (see the key derivation phase) and $\mathfrak{f}$ is a bijective function (see (ii)), we have
\begin{eqnarray}
&& \forall f \in \mathcal{F},~ \Pr(F=f) =  \Pr(V^{n_f}=\mathfrak{f}^{-1}(f)) \nonumber \\
&& \Rightarrow \eta_f - 5 N \epsilon < n_f (I(V;Y_f) - 2 \epsilon) \leq H(F) \leq \eta_f , \label{H_F}
\end{eqnarray}
Further, $B$ is selected uniformly at random from $\mathcal{B}$ of size $\eta_b$ (see (v) and the encoding phase), and so
\begin{eqnarray}
&&\forall b \in \mathcal{B},~ \Pr(B=b) =  2^{- \eta_b}~~ \Rightarrow H(B) = \eta_b. \label{H_B}
\end{eqnarray}
From (vi) and the key derivation phase, there are $2^\kappa$ choices for the key $S$; hence $H(S)\leq \kappa=({n_f}+{n_b})R_s$. For every $i\in \{1,2,\dots,{2^\kappa}\}$, the probability that $S=i$ equals to the probability that $(F,B)\in \mathcal{G}_i$. More specifically (see (\ref{eta_1-def}) and (\ref{kappa&gamma-def})),
\begin{eqnarray}
\forall i:~ \Pr(S=i) &=& \sum_{f,b \in \mathcal{G}_i} \Pr(F=f \wedge B=b)\nonumber \\
                     &\leq& 2^{\gamma} 2^{-\eta_f + 5 N \epsilon} 2^{-\eta_b} = 2^{\gamma} 2^{-\eta + 5 N \epsilon}\nonumber\\
                     &=&2^{- (\kappa - 5 N \epsilon)}\nonumber\\
\Rightarrow (n_f+n_b)(R_s- \delta) \leq \kappa - 5 N \epsilon &\leq& H(S)\leq (n_f+n_b)R_s, ~~~~\delta \geq 5 \epsilon . \label{H_S}
\end{eqnarray}

\noindent \textbf{Reliability Analysis: Proving (\ref{rel})}\\
We shall show that $S=\hat{S}$ with high probability. The encoding phase is successful with high probability: since there are $\eta_f= n_f [I(V;Y_f)+\alpha]$ sequences in $\mathcal{V}^{n_f}_{\epsilon}$, from joint-AEP, with probability arbitrarily close to 1, there exists a $V^{n_f} \in \mathcal{V}^{n_f}_{\epsilon}$ that is $\epsilon$-jointly typical with $Y_f^{n_f}$ (w.r.t. $P_{V,Y_f}$). The decoding phase includes two levels of decoding. First, Alice decodes $Y^{n_b}_b$ to $\hat{T}$ and $\hat{B}$. There are $2^{\eta_b+\eta_t}$ codewords $W_1^{n_b}$ in the codebook $\mathcal{C}_1$. From (\ref{eta_t-def}) and (\ref{eta_b-def}), we have
\begin{eqnarray*}
\eta_t+\eta_b = n_{b,2} I(W_1;Y_b) + n_{b,1} I(W_1;Y_b ) - n_b \beta < n_b I(W_1;Y_b)- 3 N\epsilon \leq n_b [I(W_1;Y_b)- 3 \epsilon].
\end{eqnarray*}
Hence, from joint-AEP, with high probability there exists a unique sequence $\hat{W}_1^{n_b}$ that is $\epsilon$-jointly typical decoding to $Y^{n_b}_b$. In the second level of decoding, Alice focuses on $\mathcal{V}_{\hat{T},\epsilon}^{n_f}$ as a codebook and looks for a unique codeword $\hat{V}^{n_f} \in \mathcal{V}_{\hat{T},\epsilon}^{n_f}$ that is $\epsilon$-jointly typical to $X^{n_f}_f$. From (i) and (iii), there are $2^{\eta_f-\eta_t}$ codewords in this codebook, and we have
\begin{eqnarray} \label{eta}
\eta_f-\eta_t    &\stackrel{(a)}{=}& n_f (I(V;Y_f)+\alpha) - n_{b,2} [I(W_{1};Y_b) -\beta] \nonumber \\
        &\stackrel{(b)}{=}& n_f(I(V;X_f,Y_f)+\alpha) - n_{b,2} I(W_{1};Y_b) + n_{b,2}\beta  \nonumber  \\
        &=& n_f I(V;X_f) + n_f(I(V;Y_f|X_f)+3\alpha) - n_{b,2} I(W_{1};Y_b) - 2n_f \alpha  + n_{b,2}\beta \nonumber \\
        &\stackrel{(c)}{<}& n_f I(V;X_f) - 3 N \epsilon \leq n_f (I(V;X_f) - 3 \epsilon) . \nonumber
\end{eqnarray}
Equality (a) follows from (\ref{eta_f-def}) and (\ref{eta_t-def}), equality (b) is due to the Markov chain $V\leftrightarrow Y_f\leftrightarrow X_f$, and inequality (c) follows from (\ref{n_{b,2}}). Hence, from joint-AEP, the appropriate $\hat{V}^{n_f}\in \mathcal{V}_{\hat{T},\epsilon}^{n_f}$ is found with high probability. The rest is key derivation which is deterministic and does not increase the error probability, i.e., the error probability at the end of the protocol is upper bounded by that of the decoding phase. This gives $\Pr(\hat{S}\neq S) < \delta$ for arbitrarily small $\delta$.
\medskip

\noindent \textbf{Secrecy Analysis: Proving (\ref{sec})}\\
We shall show that the $H(S|Z_f^{n_f},Z^{n_b}_b)$ is close to $H(S)$. We first calculate the quantities $H(T)$, $H(T_2)$, and $H(B_2)$, that are used in the sequel. From the encoding phase, $T=\mathfrak{t}(V^{n_f})$, and we have (see (i), (iii) and (\ref{eta_t-def}) and (\ref{P_V}))
\begin{eqnarray}
\forall t \in \mathcal{T},~ \Pr(T=t) &=& \sum_{v^{n_f} \in \mathcal{V}^{n_f}_{t,\epsilon}} \Pr(V^{n_f}=v^{n_f}) \nonumber \\
&\leq& 2^{\eta_f-\eta_t} 2^{-\eta_f + 5 N \epsilon} = 2^{-\eta_t + 5N \epsilon} \label{P_T} \\
\Rightarrow \eta_t- 5N \epsilon &\leq& H(T) \leq \eta_t, \label{H_T}
\end{eqnarray}
where the upper bound on $H(T)$ is due to $|\mathcal{T}|=2^{\eta_t}$. From the encoding phase $(T_2,T_1)=\mathfrak{t}_{indx}(T)$, and we have (see (iv) and (\ref{eta_t-def}) and (\ref{P_T}))
\begin{eqnarray}
\forall i \in \{1,\dots,2^{\eta_{t,2}}\},~ \Pr(T_2=i) &=& \Pr(T \in \mathcal{T}_i) = \sum_{j=1}^{\eta_{t,1}} \Pr(T=t_{i,j}) \nonumber \\
&\leq& 2^{\eta_{t,1}} 2^{-\eta_t + 5 N \epsilon} = 2^{-\eta_{t,2} + 5N \epsilon} \nonumber\\
\Rightarrow \eta_{t,2}- 5N \epsilon &\leq& H(T_2) \leq \eta_{t,2}, \label{H_T2}
\end{eqnarray}
where the upper bound follows from $|\mathcal{T}_2|=2^{\eta_{t,2}}$. Likewise $(B_2,B_1)=\mathfrak{b}_{indx}(B)$ and so, using (V) and (\ref{eta_b-def}), we have
\begin{eqnarray}
\forall i \in \{1,\dots,2^{\eta_{b,2}}\},~ \Pr(B_2=i) &=& \Pr(B \in \mathcal{B}_i) = \sum_{j=1}^{\eta_{b,1}} \Pr(B=b_{i,j}) \nonumber \\
&=& 2^{\eta_{b,1}} 2^{-\eta_b} = 2^{-\eta_{b,2}} \nonumber \\
\Rightarrow H(B_2) &=& \eta_{b,2} . \label{H_B2}
\end{eqnarray}

In Lemma \ref{lm1}, we give a lower bound for $H(S|Z_f^{n_f},Z^{n_b}_b)$. Lemma \ref{lm2} is used to show that this lower bound is arbitrarily close to $H(S)$. Finally, Corollary \ref{cor_1} uses the results of these two lemmas to prove (\ref{sec}).
\begin{lemma}\label{lm1}
Eve's uncertainty about the secret $S$, satisfies
\begin{eqnarray*}
H(S|Z_f^{n_f},Z^{n_b}_b)\geq H(S) -H(F,B|S,T_2,B_2,Z_f^{n_f},Z^{n_b}_b)-19 N \epsilon.
\end{eqnarray*}
\end{lemma}

\begin{IEEEproof}
\begin{eqnarray}
H(S|Z_f^{n_f},Z^{n_b}_b)&\geq& H(S|T_2,B_2,Z_f^{n_f},Z^{n_b}_b) \nonumber \\
                        &=& H(S,F,B|T_2,B_2,Z_f^{n_f},Z^{n_b}_b)-H(F,B|S,T_2,B_2,Z_f^{n_f},Z^{n_b}_b) \nonumber \\
                        &=& H(F,B|T_2,B_2,Z_f^{n_f},Z^{n_b}_b)-H(F,B|S,T_2,B_2,Z_f^{n_f},Z^{n_b}_b)\nonumber \\
                        &=& H(F,B|T_2,B_2)-I(F,B;Z_f^{n_f},Z^{n_b}_b|T_2,B_2)-H(F,B|S,T_2,B_2,Z_f^{n_f},Z^{n_b}_b).~~  \label{HS-bound1}
\end{eqnarray}
In (\ref{HS-bound1}), the last term  appears in the statement of Lemma \ref{lm1}, so it remains to calculate the first two terms terms. The first one is written as
{\small
\begin{eqnarray}
H(F,B|T_2,B_2) & =& H(F|T_2,B_2) + H(B|F,T_2,B_2) \stackrel{(a)}{=} H(F|T_2) + H(B|B_2)  \nonumber \\
&\stackrel{(b)}{=}& H(F) + H(B) -H(T_2) - H(B_2) \label{HS-bound1.5} \\
&\stackrel{(c)}{\geq}& \eta_f - 5N \epsilon + \eta_{b} - \eta_{t,2} - \eta_{b,2} \nonumber\\
&\stackrel{(d)}{\geq}& n_f (I(V;Y_f) - 2 \epsilon) + n_{b,1}[I(W_1;Y_b) - \beta]  - n_{b,2}I(W_2;Y_b) - n_{b,1}I(W_2;Y_b) \nonumber\\
& \stackrel{(e)}{=}& n_f I(V;X_f) + n_f I(V;Y_f|X_f) -2 n_f \epsilon + n_{b,1} I(W_{1};Y_b) - n_{b}I(W_2;Y_b) - n_{b,1} \beta \nonumber \\
& =& n_f I(V;X_f) + n_f (I(V;Y_f|X_f)+3\alpha) + n_{b,1} I(W_{1};Y_b) - n_{b}I(W_2;Y_b) - 3n_f \alpha - n_b \beta - 2 n_f \epsilon \nonumber \\
& \stackrel{(f)}{=}& n_f I(V;X_f) + n_{b,2} I(W_{1};Y_b) + n_{b,1} I(W_{1};Y_b) - n_{b}I(W_2;Y_b) - 3n_f \alpha - n_b \beta - 2 n_f \epsilon\nonumber \\
& > & n_f I(V;X_f) + n_{b} I(W_{1};Y_b) - n_{b}I(W_2;Y_b) - 14 N \epsilon\nonumber \\
& \stackrel{(g)}{=}& n_f I(V;X_f) + n_{b} I(W_{1};Y_b|W_2) - 14 N \epsilon \label{HS-bound2}
\end{eqnarray}
}
Equality (a) holds since $B_2$ and $B$ are selected independently of $T_2$ and $F$; equality (b) holds since $T_2$ and $B_2$ are deterministic functions of $F$ and $B$, respectively (see the encoding phase); inequality (c) follows from (\ref{H_F}), (\ref{H_B}), (\ref{H_T2}), and (\ref{H_B2}); equality (d) follows from (\ref{eta_f-def}), (\ref{eta_t-def}), and (\ref{eta_b-def}); equality (e) is due to the Markov chain $X_f\leftrightarrow Y_f \leftrightarrow V$ and (viii); equality (f) follows from (\ref{n_{b,2}}), and equality (g) is due to the Markov chain $W_2\leftrightarrow W_1 \leftrightarrow Y_b$. 

The second term in (\ref{HS-bound1}) is written as
{\footnotesize
\begin{eqnarray}
I(F,B;Z_f^{n_f},Z^{n_b}_b|T_2,B_2)&=& I(F,B;Z_f^{n_f}|T_2,B_2) + I(F,B;Z^{n_b}_b|Z_f^{n_f},T_2,B_2) \nonumber\\
& \stackrel{(a)}{=}& I(V^{n_f},B;Z_f^{n_f}|T_2,B_2) + I(V^{n_f},T,B;Z^{n_b}_b|Z_f^{n_f},T_2,B_2) \nonumber\\
& \stackrel{(b)}{=}& I(V^{n_f},B;Z_f^{n_f}|T_2,B_2) + I(T,B;Z^{n_b}_b|Z_f^{n_f},T_2,B_2)\nonumber \\
& \stackrel{(c)}{\leq}& I(V^{n_f},B;Z_f^{n_f}|T_2,B_2) + I(T,B;Z^{n_b}_b|T_2,B_2) \nonumber\\
& \stackrel{(d)}{\leq}& I(V^{n_f};Z_f^{n_f}) + I(T,B;Z^{n_b}_b|T_2,B_2) \nonumber\\
& =&            I(V^{n_f};Z_f^{n_f}) + \min\{ [H(T,B|T_2,B_2)], [I(T,B;Z^{n_b}_b|T_2,B_2)] \} \nonumber\\
& \stackrel{(e)}{=}&            I(V^{n_f};Z_f^{n_f}) + \min\{ [H(T|T_2)+H(B|B_2)], [H(Z^{n_b}_b|T_2,B_2)- H(Z^{n_b}_b|T,B,T_2,B_2)] \} \nonumber\\
& \stackrel{(f)}{=}&            I(V^{n_f};Z_f^{n_f}) + \min\{ [H(T)-H(T_2)+H(B)-H(B_2)],  [H(Z^{n_b}_b|T_2,B_2)- H(Z^{n_b}_b|T,B)]\} \nonumber\\
& \stackrel{(g)}{\leq}& {n_f} I(V;Z_f)  + \min\{ n_b[I(W_1;Y_b)-I(W_2;Y_b)]-5N\epsilon, [H(Z^{n_b}_b|T_2,B_2)- H(Z^{n_b}_b|T,B)]\} \nonumber \\
& \stackrel{(h)}{\leq}& {n_f} I(V;Z_f)  + \min\{ n_b[I(W_1;Y_b)-I(W_2;Y_b)]-5N\epsilon, n_b[H(Z_b|W_2)- H(Z_b|W_1)]\} \nonumber \\
& \stackrel{(i)}{\leq}& {n_f} I(V;Z_f)  + \min\{ n_bI(W_1;Y_b|W_2), n_b I(W_1;Z_b|W_2)\}-5N\epsilon \label{HS-bound3}
\end{eqnarray}
}
Inequality (a) holds because $V^{n_f}=\mathfrak{f}^{-1}(F)$ (the key derivation phase) and $T$ is a deterministic function of $V^{n_f}$ (the encoding phase); equality (b) holds because $V^{n_f} \leftrightarrow (T,B)\leftrightarrow Z_b^{n_b}$ forms a Markov chain; inequality (c) is due to the Makov chain $Z_f^{n_f}\leftrightarrow (T,B) \leftrightarrow Z_b^{n_b}$;  inequality (d) is due to the Makov chain $(B,B_2,T_2)\leftrightarrow V^{n_f} \leftrightarrow Z_f^{n_f}$;  equality (e) holds since $T_2$ and $T$ are obtained independently of $B_2$ and $B$; equality (f) holds since $T_2$ and $B_2$ are parts of $T$ and $B$, respectively; inequality (g) follows from (\ref{H_B}), (\ref{H_T}), (\ref{H_T2}), and (\ref{H_B2}); inequality (h) follows from AEP, and equality (i) is due to the Markov chain $W_2\leftrightarrow W_1 \leftrightarrow Z_b$. Applying (\ref{HS-bound2}) and (\ref{HS-bound3}) in (\ref{HS-bound1}) gives
\begin{eqnarray*}
H(S|Z_f^{n_f},Z^{n_b}_b) & >& n_f (I(V;X_f)-I(V;Z_f)) + n_b [I(W_{1};Y_b|W_{2})- I(W_{1};Z_b|W_{2}) ]_+ \\
&& - 19 N \epsilon -H(F,B|S,T_2,B_2,Z_f^{n_f},Z^{n_b}_b)\\
&=& ({n_f}+{n_b})R_s- 19 N \epsilon -H(F,B|S,T_2,B_2,Z_f^{n_f},Z^{n_b}_b)\\
&\geq& H(S) - 19 N \epsilon -H(F,B|S,T_2,B_2,Z_f^{n_f},Z^{n_b}_b),
\end{eqnarray*}
where the last inequality follows from (\ref{H_S}).
\end{IEEEproof}

\begin{lemma}\label{lm2}
$H(F,B|S,T_2,B_2,Z_f^{n_f},Z^{n_b}_b)\leq h(2\epsilon)+2\epsilon \eta$.
\end{lemma}
\begin{IEEEproof}
We shall show that the knowledge of $(S,T_2,B_2,Z_f^{n_f},Z^{n_b}_b)$ gives almost all the information about $F,B$. From (xi), knowing $S=i$ gives the partition $\mathcal{G}_{i}$ that $F,B$ belongs to; further, knowing $T_2=t_2$ and $B_2=b_2$ gives the codeword $w^{n_b}_{2,t_2,b_2} \in \mathcal{C}_2$ which is used in the encoding phase (see (xii) and (xiii)). Define the codebook  $\mathcal{C}^e_i \stackrel{\Delta}{=} \{v^{n_f},w_1^{n_{b}}:~ (\mathfrak{f}(v^{n_f}),b) \in \mathcal{G}_i,~ w_1^{n_b}=Enc(\mathfrak{t}(v^{n_f}),b),~ T_2=t_2,~ B_2=b_2\}$. Given $Z_f^{n_f},Z^{n_b}_b$, one can search all the codewords in $\mathcal{C}^e_i$ and return a unique $\hat{\hat{V}}^{n_f},\hat{\hat{W}}_1^{n_{b}} \in \mathcal{C}^e_i$ that is $(\epsilon, {n_f})$-bipartite jointly typical to $(Z^{n_f}_f,Z^{n_b}_b)$ w.r.t. $(P_{V,Z_f},P_{W_1,Z_b})$; otherwise return a NULL. From (xi), $|\mathcal{G}_{i}|=2^\gamma$, and so $|\mathcal{C}^e_{i}|=2^{\gamma-\eta_2}$, where $\eta_2=\eta_{t,2}+\eta_{b,2}$. If $\gamma-\eta_2$ is sufficiently smaller than ${n_f} I(V;Z_f) +  n_b I(W_1;Z_b)$, from joint-AEP for bipartite sequences (in Theorem \ref{theorem-AEP}), the above jointly-typical decoding will result in arbitrarily small error probability. To prove $\gamma-\eta_2$ is smaller than ${n_f} I(V;Z_f) +  n_b I(W_1;Z_b)$, we first calculate the following term.

\begin{eqnarray*}
\eta    &=& \eta_f + \eta_b \\
        &=& n_f (I(V;Y_f) + \alpha) + n_{b,1} I(W_{1};Y_b) - n_b \beta \\
        &=& n_f I(V;X_f) + n_f (I(V;Y_f|X_f)+ 3\alpha) + n_{b,1} I(W_{1};Y_b) - 2 n_f \alpha - n_b \beta \\
        &=& n_f I(V;X_f) + n_b I(W_{1};Y_b) - 3 n_f \alpha.
\end{eqnarray*}
Hence,
\begin{eqnarray*}
\gamma-\eta_2 &\stackrel{(a)}{=}& \eta-({n_f}+{n_b})R_s - \eta_{t,2} -\eta_{b,2} \\
& \stackrel{(b)}{\leq}& n_f I(V;X_f)+ n_b I(W_{1};Y_b) - 3 n_f \alpha + n_f [I(V;Z_f) - I(V;X_f)] \\
&& + n_b [I(W_1;Z_b|W_2)-I(W_1;Y_b|W_2)] - n_{b,2} I(W_2;Y_b) - n_{b,1} I(W_2;Y_b) \\
&\stackrel{(c)}{=}&   n_b I(W_{1};Y_b) - 3 n_f \alpha + {n_f} I(V;Z_f)+  n_b [I(W_1;Z_b|W_2)-I(W_1;Y_b|W_2)] - n_b I(W_2;Y_b) \\
&\stackrel{(d)}{=}&   {n_f} I(V;Z_f) + n_b I(W_1;Z_b|W_2) - 3 n_f \alpha \\
&\stackrel{(e)}{<}&   {n_f} I(V;Z_f) + n_b I(W_1;Z_b) - 9 N \epsilon.
\end{eqnarray*}

Equality (a) follows from (x) and (xi), inequality (b) follows from the definition of $R_s$ in (\ref{R_s}), equality (c) follows from (ii), equality (d) is due to the Markov chain $W_2 \leftrightarrow W_1 \leftrightarrow Y_b$, and inequality (e) is due to the Markov chain $W_2 \leftrightarrow W_1 \leftrightarrow Z_b$. From Theorem \ref{theorem-AEP} (joint-AEP for bipartite sequences), the decoding error probability becomes arbitrarily close to 0, i.e., given $(S,T_2,B_2,Z_f^{n_f},Z^{n_b}_b)$, we have $\Pr\left((\hat{\hat{V}}^{n_f},\hat{\hat{W}}_1^{n_{b}}) \neq (V^{n_f}, W^{n_b}_1)\right) < 2 \epsilon$. Let $\hat{\hat{F}}=\mathfrak{f}(\hat{\hat{V}}^{n_f})$ and $\hat{\hat{B}},\hat{\hat{T}}=Enc(\hat{\hat{W}}_1^{n_{b}})$, then we have

\begin{eqnarray*}
\Pr\left((\hat{\hat{F}},\hat{\hat{B}}) \neq (F, B)\right) < 2 \epsilon.
\end{eqnarray*}
Using Fano's inequality \cite{Ga68} results in
\begin{eqnarray*}
H(F,B|S,T,B,Z_f^{n_f},Z_b^{n_b})\leq H(F,B|\hat{\hat{F}},\hat{\hat{B}}) \leq h(2\epsilon)+2\epsilon \eta,
\end{eqnarray*}
where $h(\epsilon)=-\epsilon\log(\epsilon)-(1-\epsilon)\log(1-\epsilon)$ is the binary entropy function.
\end{IEEEproof}

\begin{corollary}\label{cor_1}
From Lemmas \ref{lm1} and \ref{lm2}, for any arbitrarily $\delta>0$, by choosing appropriately $\alpha,\beta,\epsilon >0$, Eve's uncertainty rate about $S$ is lower-bounded as
\begin{eqnarray*}
\frac{H(S|Z^{n_f}_f,Z^{n_b}_b)}{H(S)} \geq 1- \delta.
\end{eqnarray*}
\end{corollary}

\bigskip

\section{Proof of Theorem \ref{th_upperbound}, the upper bound} \label{app_B}
\begin{figure} [h]
\centering
  \includegraphics[scale=.4]{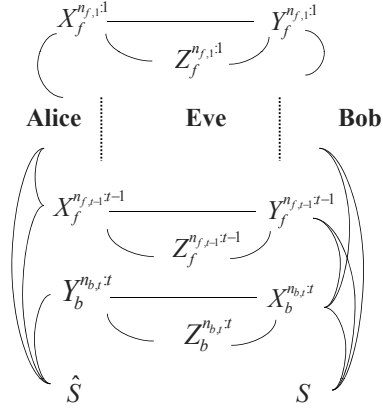}\\
  \caption{The relations between sequences of RVs in a $t$-round SKE protocol}\label{fig-upperbound}
\end{figure}

There are eight cases for a $t$-round SKE protocol, depending on the party who initiates the protocol, the one who calculates $S$, and whether $t$ is odd or even. We assume $t$ is even, Alice is the initiator, and Bob calculates $S$. The other cases can be argued similarly and lead to the same result. Alice sends $X^{n_{f,r}:r}_f$ of length $n_{f,r}$ in odd rounds $r\in \{1, 3, \dots, t-1\}$; Bob and Eve receive $Y^{n_{f,r}:r}_f$ and $Z^{n_{f,r}:r}_f$, respectively. Bob sends $X^{n_{b,r}:r}_b$ of length $n_{b,r}$ in even rounds $r\in \{2, 4, \dots, t\}$; Alice and Eve receive $Y^{n_{b,r}:r}_b$ and $Z^{n_{b,r}:r}_b$, respectively. Note that the forward and the backward channels are assumed to be used $n_f$ and $n_b$ times, respectively, and so
\begin{IEEEeqnarray*}{l}
n_f = \sum_{r\in \{1,3,\dots, t-1\}} n_{f,r},~ \mbox{ and }~
n_b = \sum_{r\in \{2,4,\dots, t\}} n_{b,r}.
\end{IEEEeqnarray*}
We denote views of Alice, Bob, and Eve at the end of round $r$, by $V^{:r}_A$, $V^{:r}_B$, and $V^{:r}_E$, respectively. For instance $V^{:r}_A$ is
\begin{IEEEeqnarray*}{lll}
V^{:r}_A=\left( ||_{(odd) i< r} \left[X^{n_{f,i}:i}_f \right]\right) ~||~ \left(||_{even: i\leq r}\left[Y^{n_{b,i}:i}_b \right]\right).
\end{IEEEeqnarray*}
$V^{:r}_B$ and $V^{:r}_E$ can be presented similarly. Fig. \ref{fig-upperbound} illustrates the relationships between the sequences of RVs (and the keys), where two sequences are connected by an edge if (i) they belong to input/outputs of the same DMBC, or (ii) one is computed from the other by Alice or Bob, using a (possibly randomized) function.

For an even $r$, at the end of round $r-1$, Bob computes the sequence $X^{n_{b,r}:r}_b$ using his view, $V^{:r-1}_B$, through a (possibly randomized) function $\phi_r(R,V^{:r-1}_B)$ where the randomness $R$ is independent of other parties' views $V^{:r-1}_A$ and $V^{:r-1}_E$. He sends this sequence in round $r$, where the received sequences $Y^{n_{b,r}:r}_f$ and $Z^{n_{b,r}:r}_f$ are determined from $X^{n_{b,r}:r}_b$ through the backward channel transition matrix that is independently of the the views in round $r-1$. Accordingly,
\begin{IEEEeqnarray*}{l}
(V^{:r-1}_A, V^{:r-1}_E)  \leftrightarrow V^{:r-1}_B \leftrightarrow X^{n_{b,r}:r}_b \leftrightarrow (Y^{n_{b,r}:r}_b,Z^{n_{b,r}:r}_b)
\end{IEEEeqnarray*}
forms a Markov chain, from which we derive the following four Markov chains, specifically used in the sequel,
\begin{IEEEeqnarray}{l}
V^{:r-1}_B\leftrightarrow X^{n_{b,r}:r}_b \leftrightarrow Y^{n_{b,r}:r}_b, \IEEEyessubnumber \label{markov-rounf-r-a} \\
V^{:r-1}_E\leftrightarrow X^{n_{b,r}:r}_b \leftrightarrow Y^{n_{b,r}:r}_b, \IEEEyessubnumber \label{markov-rounf-r-b}\\
(Y^{n_{b,r}:r}_b,Z^{n_{b,r}:r}_b) \leftrightarrow V^{:r-1}_B \leftrightarrow V^{:r-1}_A, \IEEEyessubnumber \label{markov-rounf-r-c}\\
X^{n_{b,r}:r}_b \leftrightarrow V^{:r-1}_B \leftrightarrow V^{:r-1}_A. \IEEEyessubnumber \label{markov-rounf-r-d}
\end{IEEEeqnarray}
By symmetry, one can show Markov chains between variables when $r$ is odd. The views of the parties at the end of the protocol are then $View_A=V^{:t}_A$, $View_B=V^{:t}_B$, and $View_E=V^{:t}_E$. Bob computes the key $S \in \mathcal{S}$ as a function of $V^{:t}_B$ and Alice computes $\hat{S}\in \mathcal{S}$ as a function of $V^{:t}_A$.
Note that the rate $R_{sk}$ for an arbitrarily small $\delta>0$ is achievable if (\ref{SKE-Eqs}) is satisfied. Using Fano's inequality for (\ref{SKE-rel}), we have
\begin{eqnarray}
H(S|View_A) \leq H(S|\hat{S}) < h(\delta) + \delta H(S) , \label{H(S|S')}
\end{eqnarray}
Furthermore, (\ref{SKE-sec}) gives
\begin{eqnarray}\label{I(S;Z)}
I(S;View_E)=H(S)-H(S|View_E) \leq \delta H(S).
\end{eqnarray}
For given $n_f$ and $n_b$, $H(S)$ is upper bounded as
\begin{IEEEeqnarray}{lll} \label{H(S)}
H(S)&=&    I(S;V^{:t}_A) + H(S|V^{:t}_A) \nonumber \\
    &\stackrel{(a)}{\leq}&  I(S;V^{:t}_A) + h(\delta) + \delta H(S) - I(S;V^{:t}_E) + I(S;V^{:t}_E) \nonumber\\
    &\stackrel{(b)}{\leq}&  I(S;V^{:t}_A) - I(S;V^{:t}_E) + h(\delta) + 2 \delta H(S) \nonumber \\
    &\leq&  I(S;V^{:t}_A | V^{:t}_E) + h(\delta) + 2 \delta H(S) \label{H(S)-1} \\
    &\stackrel{(c)}{\leq}&  I(V^{:t}_B;V^{:t}_A | V^{:t}_E) + h(\delta) + 2 \delta H(S)\nonumber  \\
    && \Rightarrow H(S) \leq \frac{1}{1-2\delta} [I(V^{:t}_B;V^{:t}_A | V^{:t}_E) + h(\delta)].
\end{IEEEeqnarray}
Inequalities (a) and (b) follow from (\ref{H(S|S')}) and (\ref{I(S;Z)}), respectively, and inequality (c) follows from the Markov chain $S \leftrightarrow V^{:t}_B \leftrightarrow V^{:t}_A$. The first term in (\ref{H(S)}) is written as follows
\begin{IEEEeqnarray}{l} \label{bound-round:t}
I(V^{:t}_B;V^{:t}_A | V^{:t}_E) \nonumber \\
\tab = I(V^{:t}_B; Y^{n_{b,t}:t}_b |V^{:t}_E )+I(V^{:t}_B; V^{:t-1}_A|V^{:t}_E,Y^{n_{b,t}:t}_b)\nonumber\\
\tab \stackrel{(a)}{=} I(X^{n_{b,t}:t}_b; Y^{n_{b,t}:t}_b |V^{:t}_E )+I(V^{:t}_B; V^{:t-1}_A|V^{:t}_E,Y^{n_{b,t}:t}_b)\nonumber\\
\tab \stackrel{(b)}{\leq} I(X^{n_{b,t}:t}_b; Y^{n_{b,t}:t}_b | Z^{n_{b,t}:t}_b)+I(V^{:t}_B; V^{:t-1}_A|V^{:t}_E,Y^{n_{b,t}:t}_b)\nonumber\\
\tab \stackrel{(c)}{\leq} I(X^{n_{b,t}:t}_b; Y^{n_{b,t}:t}_b | Z^{n_{b,t}:t}_b)+I(V^{:t}_B; V^{:t-1}_A|V^{:t-1}_E) \nonumber\\
\tab \stackrel{(d)}{=} I(X^{n_{b,t}:t}_b; Y^{n_{b,t}:t}_b | Z^{n_{b,t}:t}_b)+I(V^{:t-1}_B; V^{:t-1}_A|V^{:t-1}_E). ~~~
\end{IEEEeqnarray}
(In)equalities (a)-(d) follow from the Markov chains (\ref{markov-rounf-r-a})-(\ref{markov-rounf-r-d}), respectively, when $r=t$. By symmetry, one can write the second term in (\ref{bound-round:t}) as
\begin{IEEEeqnarray}{l} \label{bound-round:t-1}
I(V^{:t-1}_B; V^{:t-1}_A|V^{:t-1}_E) \leq I(Y^{n_{f,t-1}:t-1}_f;X^{n_{f,t-1}:t-1}_f |Z^{n_{f,t-1}:t-1}_f) + I(V^{:t-2}_B;V^{:t-2}_A | V^{:t-2}_E).\tab
\end{IEEEeqnarray}
Repeating the steps in (\ref{bound-round:t}) and (\ref{bound-round:t-1}) $t/2$ times, we arrive at
\begin{IEEEeqnarray}{l} \label{bound-round:0}
I(V^{:t}_B;V^{:t}_A | V^{:t}_E) \leq \sum_{(odd) r< t} I(Y^{n_{f,t}:r}_f;X^{n_{f,t}:r}_f |Z^{n_{f,t}:r}_f) + \sum_{(even) r\leq t} I(X^{n_{b,t}:r}_b; Y^{n_{b,t}:r}_b | Z^{n_{b,t}:r}_b).\tab
\end{IEEEeqnarray}
For an odd (resp. even) $r$, define $X^{:r}_f$ (resp. $X^{:r}_b$) such that
\begin{IEEEeqnarray*}{l}\label{input-dist}
P_{X^{:r}_f}=\frac{1}{n_{f,r}}\sum_{i=1}^{n_{f,r}} P_{X^{:r}_{f,i}}, \tab (\mbox{resp. } P_{X^{:r}_b}=\frac{1}{n_{b,r}} \sum_{i=1}^{n_{b,r}} P_{X^{:r}_{b,i}}~).
\end{IEEEeqnarray*}
Obtain $Y^{:r}_f,Z^{:r}_f$ (resp. $Y^{:r}_b,Z^{:r}_b$) from the 2DMBC conditional distributions. We choose the RVs $X_f, Y_f, Z_f$ and $X_b, Y_b, Z_b$ that correspond to the 2DMBC distributions ($P_{Y_f,Z_f|X_f}$ and $P_{Y_b,Z_b|X_b}$), and $X_f$ and $X_b$ are selected to satisfy
\begin{IEEEeqnarray*}{l}\label{input-dist}
I(X_f;Y_f|Z_f)=\max_{(odd)r<t} [I(X^{:r}_f;Y^{:r}_f|Z^{:r}_f)], \tab
I(X_b;Y_b|Z_b)=\max_{(even)r<t} [I(X^{:r}_b;Y^{:r}_b|Z^{:r}_b)],
\end{IEEEeqnarray*}
respectively. We continue (\ref{bound-round:0}) as
\begin{IEEEeqnarray}{l} \label{bound-round:total}
I(V^{:t}_B;V^{:t}_A | V^{:t}_E) \nonumber \\
~\stackrel{(a)}{\leq} \sum_{(odd) r < t} n_{f,r} I(Y_f;X_f |Z_f) + \sum_{even: r\leq t} n_{b,r} I(X_b;Y_b|Z_b) \nonumber\\
~=n_f  I(X_f;Y_f|Z_f)+ n_b I(X_b;Y_b | Z_b).
\end{IEEEeqnarray}

Inequality (a) follows from Jensen's inequality since $I(X_f;Y_f|Z_f)$ and $I(X_b;Y_b|Z_b)$ are concave functions of $P_{X_f}$ and $P_{X_b}$, respectively (see e.g., \cite[Appendix-I]{Kh08-2}). We have shown that for any SKE protocol, there exist RVs for which (\ref{bound-round:total}) holds.

Using (\ref{SKE-rand}), (\ref{H(S)}) and (\ref{bound-round:total}), we have the following upper bound on $R_{sk}$
\begin{IEEEeqnarray*} {lll}
 R_{sk} & <& \frac{1}{n_f+n_b} H(S) + \delta \\
                & <& \frac{n_f I(X_f;Y_f | Z_f) + n_b I(X_b;Y_b | Z_b)+h(\delta)}{(1-2\delta)(n_f+n_b)} + \delta \\
                &\leq& \max \{I(X_f;Y_f | Z_f), I(X_b;Y_b | Z_b)\},
\end{IEEEeqnarray*}
where the last inequality follows from the fact that $\delta$ is arbitrarily small. This proves the upper bound in (\ref{upper-bound}).
\bigskip

\section{Proof of Theorem \ref{th_degraded}, degraded 2DMBCs}\label{app_C}
\begin{lemma}\label{lemma-degraded}
For the degraded DMBC as defined in Definition \ref{deg-DMBC}, we have $I(X;Y|Z) \leq I(X_O;Y_O|Z_O)$.
\end{lemma}
\begin{IEEEproof}
\begin{IEEEeqnarray*}{l}
I(X;Y|Z) = I(X_O,X_R ;Y_O,Y_R|Z_O,Z_R) \\
= I(X_O;Y_O,Y_R|Z_O,Z_R) + I(X_R ;Y_O,Y_R|Z_O,Z_R,X_O) \\
\stackrel{(a)}{=} I(X_O;Y_O|Z_O,Z_R) + I(X_R ;Y_R|Z_O,Z_R,X_O) \\
\stackrel{(b)}{=} I(X_O;Y_O|Z_O,Z_R) \stackrel{(c)}{\leq} I(X_O;Y_O|Z_O).
\end{IEEEeqnarray*}
Equalities (a) is due to the Markov chains $X_O \leftrightarrow Z_R \leftrightarrow Y_R$ and $X_R \leftrightarrow X_O \leftrightarrow Y_O$, equality (b) is due to $X_R \leftrightarrow Z_R \leftrightarrow Y_R$, and equality (c) is due to $Z_R \leftrightarrow X_O \leftrightarrow Y_O$.
\end{IEEEproof}

$C^{d-2DMBC}_{sk}$ is upper bounded as (see Theorem \ref{th_upperbound})
\begin{IEEEeqnarray}{l}
C^{d-2DMBC}_{sk} \leq \max_{P_{X_f},P_{X_b}} \{ I(X_{f};Y_{f}|Z_{f}) , I(X_{b};Y_{b}|Z_{b}) \} \nonumber \\
~ \leq \max_{P_{X_f},P_{X_b}} \{ I(X_{f,O};Y_{f,O}|Z_{f,O}) , I(X_{b,O};Y_{b,O}|Z_{b,O}) \},\tab \label{C^d_sk-upper}
\end{IEEEeqnarray}
where the last inequality follows from Lemma \ref{lemma-degraded}. On the other hand, the lower bounded in (\ref{C^SC}) holds for $C^{d-2DMBC}_{sk}$. Starting from (\ref{L_A}), we write $L_A$ as
\begin{IEEEeqnarray*}{l}
L_A \stackrel{(a)}{\geq} \max_{n_f,n_b, P_{X_f},P_{X_b}} \left[\frac{n_b [I(X_{b,O};Y_b)-I(X_{b,O};Z_b)]_+}{n_f+n_b} \right] \\
~~ \stackrel{(b)}{\geq} \max_{P_{X_{b,O}}} [I(X_{b,O};Y_b)-I(X_{b,O};Z_b)]_+  \\
~~ \stackrel{(c)}{=} \max_{P_{X_{b,O}}} [I(X_{b,O};Y_{b,O})-I(X_{b,O};Z_{b,O})]_+  \\
~~ = \max_{P_{X_{b,O}}} [I(X_{b,O};Y_{b,O}|Z_{b,O}) ]. \IEEEyesnumber \label{L_A-lower}
\end{IEEEeqnarray*}
Inequality (a) follows from choosing $V_f=0$, $W_{2,b}=0$, and $W_{1,b}=X_{b,O}$. Since the argument to be maximized in the right hand of inequality (a) is independent of $P_{X_f}$, we remove $P_{X_f}$ from the expression. Inequality (b) is obtained by choosing $n_b$ sufficiently larger than $n_f$ and letting $X_{b,R}$ have a constant value. Equality (c) holds since $X_{b,R}$, and hence $Y_{b,R}$ and $Z_{b,R}$, are independent of $X_{b,O}$. By symmetry, one can show that
\begin{IEEEeqnarray}{l}
L_B \geq \max_{P_{X_{f,O}}} [I(X_{f,O};Y_{f,O}|Z_{f,O}) ]. \label{L_B-lower}
\end{IEEEeqnarray}
Combining (\ref{C^d_sk-upper})-(\ref{L_B-lower}) proves the theorem.

\bigskip
\section{Proof of Theorem \ref{theorem-AEP}, Joint-AEP for bipartite sequences (in Appendix \ref{app_A})}\label{app_E}
\emph{Part 1) To prove $\Pr((X^N,Y^N)\in A_\epsilon^{(N,n)})\rightarrow 1$}\\
We shall show that with high probability $X^N$ and $Y^N$ are $(\epsilon,n)$-bipartite typical sequences as in (\ref{def-x-typical}) and $(X^N,Y^N)$ satisfy (\ref{def-xy-typical}) in Definition \ref{defbip_joi_typ}. For large enough $n$ and $d$, by the weak law of large numbers, we have
\begin{eqnarray*}
&-\frac{1}{n}\log P_U(U^n)\rightarrow -E[\log P_U(U)]=H(U) \mbox{ in probability }\\
&\Rightarrow \exists n_1:~ \forall n>n_1, \Pr(|-\frac{1}{n}\log P_{U}(U^n)-H(U)|>\epsilon)<\frac{\epsilon}{6},
\end{eqnarray*}
Similarly, we can conclude the following for the other parts of the sequences.
\begin{eqnarray*}
&\exists d_1:~ \forall d>d_1, \Pr(|-\frac{1}{d}\log P_T(T^d)-H(T)|>\epsilon)<\frac{\epsilon}{6},\\
&\exists n_2:~ \forall n>n_2, \Pr(|-\frac{1}{n}\log P_{U'}(U'^n)-H(U')|>\epsilon)<\frac{\epsilon}{6},\\
&\exists d_2:~ \forall d>d_2, \Pr(|-\frac{1}{d}\log P_{T'}(T'^d)-H(T')|>\epsilon)<\frac{\epsilon}{6}.\\
\end{eqnarray*}
Since these sequences are i.i.d., we have
\begin{eqnarray*}
&\log P(X^N)=\log P_{U}(U^n) + \log P_{T}(T^d),\\
&\log P(Y^N)=\log P_{U'}(U'^n) + \log P_{T'}(T'^d),
\end{eqnarray*}
which finally results in
\begin{eqnarray}\label{X-typical}
&\forall n>n_1,\forall d>d_1, \Pr(|-\frac{1}{N}\log P(X^N)-\frac{nH(U)+dH(T)}{N}|>\epsilon)<\frac{\epsilon}{3},\\\label{Y-typical}
&\forall n>n_2,\forall d>d_2, \Pr(|-\frac{1}{N}\log P(Y^N)-\frac{nH(U')+dH(T')}{N}|>\epsilon)<\frac{\epsilon}{3}.
\end{eqnarray}
The same approach results in the following relations for the joint distribution,
\begin{eqnarray}
&\exists n_3:~ \forall n>n_3, \Pr(|-\frac{1}{d}\log P_{T,T'}(T^d,T'^d)-H(T,T')|>\epsilon)<\frac{\epsilon}{6},\nonumber \\
&\exists d_3:~ \forall d>d_3, \Pr(|-\frac{1}{n}\log P_{U,U'}(U^n,U'^n)-H(U,U')|>\epsilon)<\frac{\epsilon}{6}, \nonumber \\\label{XY-typical}
&\Rightarrow \forall n>n_3,\forall d>d_3, \Pr(|-\frac{1}{N}\log P(X^N,Y^N)-\frac{nH(U,U')+dH(T,T')}{N}|>\epsilon)<\frac{\epsilon}{3}.
\end{eqnarray}
By choosing $n>\max\{n_1,n_2,n_3\}$ and $d>\max\{d_1,d_2,d_3\}$, (\ref{X-typical}), (\ref{Y-typical}), and (\ref{XY-typical}) are satisfied. The probability union bound (over these three equations) states that $(X^N,Y^N)\notin A_\epsilon^{(N,n)}$ holds with probability less than $\epsilon$, i.e., $\Pr((X^N,Y^N)\in A_\epsilon^{(N,n)})\geq 1-\epsilon$. This proves the first part of the theorem.
\\
\\
\emph{Part 2) To prove $(1-\epsilon)2^{nH(U,U')+dH(T,T')-N\epsilon} \leq |A_\epsilon^{(N,n)}|\leq 2^{nH(U,U')+dH(T,T')+N\epsilon}$}\\
\begin{center}
$\displaystyle 1=\sum P(x^N,y^N)\geq \sum_{A_\epsilon^{(N,n)}} P(x^N,y^N)$
$\displaystyle \stackrel{(a)}{\geq} |A_\epsilon^{(N,n)}| 2^{-(nH(U,U')+dH(T,T')+N\epsilon)}$\\
$\displaystyle \Rightarrow |A_\epsilon^{(N,n)}| \leq 2^{nH(U,U')+dH(T,T')+N\epsilon}$,
\end{center}
and
\begin{center}
$\displaystyle 1-\epsilon \leq \sum_{A_\epsilon^{(N,n)}} P(x^N,y^N)$
$\displaystyle \stackrel{(b)}{\leq} |A_\epsilon^{(N,n)}| 2^{-nH(U,U')-dH(T,T')+N\epsilon}$\\
$\displaystyle \Rightarrow |A_\epsilon^{(N,n)}| \geq (1-\epsilon) 2^{nH(U,U')+dH(T,T')-N\epsilon}$.
\end{center}
Both inequalities (a) and (b) follow (\ref{XY-typical}).
\\
\\
\noindent \emph{Part 3) To prove $(1-\epsilon)2^{-nI(U;U')-dI(T;T')-3N\epsilon} \leq \Pr((\tilde{X}^N,\tilde{Y}^N)\in A_\epsilon^{(N,n)})\leq 2^{-nI(U;U')-dI(T;T')+3N\epsilon}$}\\
Note that $\tilde{X}^N$ and $\tilde{Y}^N$ are independent and $Pr(\tilde{X}^N=x^N,\tilde{Y}^N=y^N)=P(x^N)P(y^N)$. Using (\ref{X-typical}), (\ref{Y-typical}), and (\ref{XY-typical}), we have
\begin{center}
$\displaystyle \Pr((\tilde{X}^N,\tilde{Y}^N)\in A_\epsilon^{(N,n)})= \sum_{A_\epsilon^{(N,n)}} P(x^N)P(y^N)$
$\displaystyle \leq \left(2^{nH(U,U')+dH(T,T')+N\epsilon}\right) \left(2^{-nH(U)-dH(T)+N\epsilon}\right) \left(2^{-nH(U')-dH(T')+N\epsilon}\right)$\\
$\displaystyle = 2^{-nI(U;U')-dI(T;T')+3N\epsilon}$,
\end{center}
and
\begin{center}
$\displaystyle \Pr((\tilde{X}^N,\tilde{Y}^N)\in A_\epsilon^{(N,n)})= \sum_{A_\epsilon^{(N,n)}} P(x^N)P(y^N)$\\
$\displaystyle \geq (1-\epsilon) \left(2^{nH(U,U')+dH(T,T')-N\epsilon}\right) \left(2^{-nH(U)-dH(T)-N\epsilon}\right) \left(2^{-nH(U')-dH(T')-N\epsilon}\right)$\\
$\displaystyle =(1-\epsilon) 2^{-nI(U;U')-dI(T;T')-3N\epsilon}$.
\end{center}

\end{document}